\documentclass[review,authoryear]{elsarticle}

\usepackage{url}  
\usepackage{rotating}

\usepackage{amsmath}

\begin{document}

\title{Fluid-particle flow and validation using two-way-coupled
mesoscale SPH-DEM}

\author[twente]{Martin Robinson\corref{cor1}}
\ead{m.j.robinson@utwente.nl}
\author[twente]{Stefan Luding}
\ead{s.luding@utwente.nl}
\author[nestle]{Marco Ramaioli}
\ead{Marco.Ramaioli@rdls.nestle.com}
\cortext[cor1]{Corresponding author}

\address[twente]{Multiscale Mechanics, University of Twente, Enschede, Netherlands}
\address[nestle]{Nestl\'{e} Research Center, Lausanne, Switzerland}


\begin{abstract} 

First, a meshless simulation method is presented for multiphase 
fluid-particle flows with a two-way coupled Smoothed Particle 
Hydrodynamics (SPH) for the fluid and the Discrete Element Method (DEM) for the 
solid phase.  The unresolved fluid model, based on the locally 
averaged Navier Stokes equations, is expected to be considerably
faster than fully resolved models. Furthermore, in contrast to similar 
mesh-based Discrete Particle Methods (DPMs), our
purely particle-based method enjoys the flexibility that comes from the
lack of a prescribed mesh. It is suitable for problems such as free surface
flow or flow around complex, moving and/or intermeshed geometries
and is applicable to both dilute and dense particle flows. 

Second, a comprehensive validation procedure for fluid-particle simulations is
presented and applied here to the SPH-DEM method, using simulations of single
and multiple particle sedimentation in a 3D fluid column and comparison with
analytical models. Millimetre-sized particles are used along with three
different test fluids: air, water and a water-glycerol solution. The velocity
evolution for a single particle compares well (less than 1\% error) with the
analytical solution as long as the fluid resolution is coarser than two times
the particle diameter. Two more complex multiple particle sedimentation problems
(sedimentation of a homogeneous porous block and an inhomogeneous Rayleigh
Taylor instability) are also reproduced well for porosities $0.6 \le \epsilon
\le 1.0$, although care should be taken in the presence of high porosity
gradients.

Overall the SPH-DEM method successfully reproduces quantitatively the expected
behaviour in the test cases, and promises to be a flexible and accurate tool for
other, realistic fluid-particle system simulations.

\end{abstract}



\maketitle

\section{Introduction}

Fluid-particle systems are ubiquitous in nature and industry. Sediment transport
and erosion are important in many environmental studies and the interaction
between particles and interstitial fluid affects the rheology of avalanches,
slurry flows and soils. In industry, the efficiency of a fluidised bed process
(e.g. Fluidized Catalytic Cracking) is completely determined by the complex
two-way interaction between the injected gas flow and the solid granular
material. Also, the dispersion of solid particles in a fluid is of broad
industrial relevance to the food, chemical and painting industries, which
involves in most cases three phases: a granular medium, the air initially
present in its pores and an injected liquid.

The length-scale of interest determines the method of simulation for
fluid-particle systems. For very small scale processes it is feasible to fully
resolve the interstitial fluid between the particles (see
\citet{zhu99pore,pereira10sph,potapov01liquid,wachmann98collective} for a few
examples of particle or pore-scale simulations).  However, for many applications
the dynamics of interest occur over length scales much larger than the particle
diameter and the computational effort required to resolve the pore-scale is too
great. It then becomes necessary to use unresolved, or mesoscale, fluid
simulations. This mesoscale is the focus of this paper and the domain of
applicability for the SPH-DEM method. At even larger length scales of interest
(macroscale) it becomes infeasible to model the granular material as a discrete
collection of grains and instead a continuum model is used in a two-fluid model.
However, it must be noted that while this approach might be computationally
necessary in many cases, it can fail for some systems involving dense granular
flow, where existing continuum models for granular material do not adequately
reproduce important material properties such as anisotropy, history dependency,
jamming and segregation.

Fluid-particle simulations at the mesoscale are often given the term Discrete
Particle Models (DPM). These models fully resolve the individual solid particles
using a Lagrangian model for the solid phase. The fluid phase does not resolve
the interstitial fluid, but instead models the locally averaged Navier-Stokes
equations and is coupled to the solid particles using appropriate drag closures.
Most of the prior work on DPMs have been done using grid-based methods for the
fluid phase, and a few relevant examples can be seen in the papers by
\citet{tsuji93discrete}, \citet{xu97numerical,xu00numerical},
\citet{hoomans96discrete,hoomans00granular} or \citet{chu08numerical}.

Fixed pore flow simulations (where the geometry of the solid particles is
unchanging over time) using SPH for the (unresolved) fluid phase have been
described by \citet{li07saturated} and \citet{jiang07mesoscale}, but these do
not allow for the motion and collision of solid grains.
\citet{cleary06prediction} and \citet{fernandez11using} simulate slurry flow at
the mesoscale using SPH and DEM in SAG mills and through industrial banana
screens, but only perform a one-way coupling between the solid and fluid phases.

The DPM model presented in this paper is based on the locally averaged
Navier-Stokes (AVNS) equations that were first derived by Anderson and Jackson
in the sixties \citep{anderson67fluid}, and have been used with great success to
model the complex fluid-particle interactions occurring in industrial fluidized
beds \citep{deen07review}. Anderson and Jackson defined a smoothing operator
identical to that used in SPH and used it to reformulate the NS equations in
terms of smoothed variables and a local porosity field (porosity refers to the
fraction of fluid in a given volume). Given its theoretical basis in kernel
interpolation, it is natural to consider the use of the SPH method to solve the
AVNS equations, coupled with a DEM model for the solid phase.

The coupling of SPH and DEM results in a purely particle-based solution method
and therefore enjoys the flexibility that is inherent in these methods.  This
is the primary advantage of this method over existing grid-based DPMs. In
particular, the model described in this paper is well suited for
applications involving a free surface, including (but not limited to) debris
flows, avalanches, landslides, sediment transport or erosion in rivers and
beaches, slurry transport in industrial processes (e.g.  SAG mills) and
liquid-powder dispersion and mixing in the food processing industry.

Another advantage of using a DPM, or mesoscale simulation, is of course the
reduced computational requirements over a fully resolved simulation. We have
found that in general a fluid resolution of $h = 2d$ minimises the error in the
SPH-DEM method, where $d$ is the solid particle diameter. For a fully resolved
simulation the interstitial fluid must be resolved, and therefore the fluid
resolution would need to be at least $h = 0.2d$, which scales the number of
computational nodes (for the fluid) by a factor of 1000.

\begin{figure}[htp]
\centering
\includegraphics[width=0.6\textwidth]{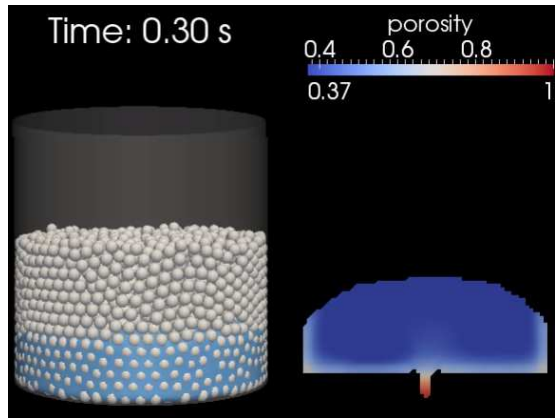}
\caption{Example of a two-phase SPH-DEM simulation of a water jet (bottom)
injected into a granular bed. On the left is shown the cell
geometry along with spheres representing the solid grains and the water surface
(coloured blue). On the right is shown the porosity profile along the plane
given by x=0. Black indicates no fluid. }
\label{fig:dispersionExample}
\end{figure}

Figure \ref{fig:dispersionExample} shows a SPH-DEM simulation applied to a
liquid-powder mixing problem in the food processing industry, taken from a
simulation of a water jet injected in a granular bed whose pores are initially
filled with air. To predict the shape of the front correctly, one has to
consider the free surface and the absence of dissipation on the air side, both
in the SPH-DEM model. Even more complex (realistic) injection geometries are
easily incorporated into the simulation with no additional effort. Moreover,
using DEM enables studying the effect (on the initial liquid front propagation)
of packing and top surface inhomogeneities that can be generated during pouring,
unlike simpler ``porous media"-like approaches.
Polydispersity can also be included by altering the radius of the simulated
grains and using a suitable drag term (e.g., see \citet{hoef05lattice})

Sections \ref{sec:GovEq}-\ref{sec:SPHDEMModel} describe the AVNS equations and
the SPH and DEM models for the fluid and solid phases and the coupling between
them. Section \ref{sec:ValidationTestCases} introduces the test cases, Section
\ref{sec:SPS} describes the results for the Single Particle Sedimentation test
case, Section \ref{sec:CPB} the results for Multiple Particle Sedimentation and
Section \ref{sec:RTI} describes the inhomogeneous Rayleigh Taylor Instability
test using solid particles sedimenting into a clear fluid.


\section{Governing Equations}\label{sec:GovEq}
\subsection{The Locally Averaged Navier-Stokes Equations} \label{sec:AVNS}

Here we describe the governing equations for the fluid phase, the locally
averaged Navier-Stokes equations derived by \citet{anderson67fluid}.  Anderson and Jackson defined a local averaging based
on a radial smoothing function $g(r)$.  The function $g(r)$ is greater than
zero for all $r$ and decreases monotonically with increasing $r$, it possesses
derivatives $g^{n}(r)$ of all orders and is normalised so that $\int g(r)dV =
1$.

The local average of any field $a'$ defined over the fluid domain can be obtained by convolution
with the smoothing function

\begin{equation}\label{eq:AVNSvariable}
\epsilon(x) a(x) = \int_{V_{f}} a'(y)g(x-y)dV_y,
\end{equation}

where $x$ and $y$ are position coordinates (here one dimensional for
simplicity). The integral is taken over the volume of interstitial fluid $V_f$ and
$\epsilon(x)$ is the porosity.

\begin{equation}\label{eq:AVNSporosity}
\epsilon(x) = 1 - \int_{V_{s}} g(x-y)dV_y,
\end{equation}

where $V_s$ is the volume of the solid particles.

In a similar fashion, the local average of any field $a'(x)$ defined over the solid domain is given by 

\begin{equation}\label{eq:AVNSvariable2}
(1 - \epsilon(x)) a(x) = \int_{V_{s}} a'(y)g(x-y)dV_y,
\end{equation}

where the integral is taken over the volume of the solid particles.

Applying this averaging method to the Navier-Stokes equations, \citet{anderson67fluid} derived the following continuity equation in
terms of locally averaged variables

\begin{equation}\label{eq:aveContinuity}
\frac{\partial (\epsilon \rho_f)}{\partial t} + \nabla \cdot (\epsilon \rho_f \mathbf{u} ) = 0,
\end{equation}

where $\rho_f$ is the fluid mass density and $\mathbf{u}$ is the fluid velocity.

The corresponding momentum equation is

\begin{equation}\label{eq:aveMomentum2}
\epsilon \rho_f \left ( \frac{\partial \mathbf{u}}{\partial t} + \mathbf{u} \cdot \nabla \mathbf{u} \right ) 
         = -\nabla P + \nabla \cdot \boldsymbol{\tau} - n \mathbf{f} + \epsilon \rho_f \mathbf{g},
\end{equation}

where $P$ is the fluid pressure, $\boldsymbol{\tau}$ is the viscous stress tensor and $n \mathbf{f}$ is
the fluid-particle coupling term. We use a Newtonian fluid where $\boldsymbol{\tau}=\mu \nabla \cdot \mathbf{u}$. We neglect Reynolds-like terms and do not consider turbulent flow. The coefficient for the coupling term $n$ is
the local average of the number of particles per unit volume and $\mathbf{f}$
is the local mean value of the force exerted on the particles by the fluid. This force includes all effects, both static and dynamic, of the particles on the fluid, the details of which can be seen in Eq.\ (\ref{Eq:demCouplingForce}).

\subsection{Smoothed Particle Hydrodynamics}

Smoothed Particle Hydrodynamics \citep{gingold77smoothed, lucy77numerical,
monaghan05SPH} is a Lagrangian scheme, whereby the fluid is discretised into
``particles'' that move with the local fluid velocity. Each particle is assigned a mass
and can be thought of as the same volume of fluid over time. The fluid
variables and the equations of fluid dynamics are interpolated over each
particle and its nearest neighbours using a smoothing kernel $W(r,h)$, where $h$ is the smoothing length scale. Like 
$g(r)$ in the AVNS equations, the SPH kernel is a radial
function that decreases monotonically and is normalised so that $\int W(r,h)dV = 1$.

Unlike $g(r)$ and to reduce the computational burden of the method, the SPH
kernel is normally defined with a compact support and a finite number of
derivatives.  

In SPH, a fluid variable $A(\mathbf{r})$ (such as momentum or density) is
interpolated using the kernel $W$

\begin{equation}\label{Eq:integralInterpolant}
A(\mathbf{r}) = \int A(\mathbf{r'})W(\mathbf{r}-\mathbf{r'},h)d\mathbf{r'}.
\end{equation}

To apply this to the discrete SPH particles, the integral is replaced by a sum
over all particles, commonly known as the \emph{summation interpolant}. To
estimate the value of the function $A$ at the location of particle $a$ (denoted
as $A_a$), the summation interpolant becomes

\begin{equation}\label{Eq:summationInterpolant}
A_a = \sum_b m_b \frac{A_b}{\rho_b} W_{ab}(h_a),
\end{equation}
where $m_b$ and $\rho_b$ are the mass and density of particle $b$. The volume
element $d\mathbf{r'}$ of Eq.\ (\ref{Eq:integralInterpolant}) has been replaced
by the volume of particle $b$ (approximated by $\frac{m_b}{\rho_b}$),
equivalent to the normal trapezoidal quadrature rule.  The kernel function is
denoted by $W_{ab}(h) = W(\mathbf{r}_a-\mathbf{r}_b,h)$. The dependence of the
kernel on the difference in particle positions is not explicitly stated for
readability. Due to the limited support of W, particle neighbourhood search
methods as standard in SPH or DEM can be applied to optimize the summation in Eq.
(\ref{Eq:integralInterpolant}). 

The accuracy of the SPH interpolant depends on the particle positions within
the radius of the kernel. If there is not a homogeneous distribution
of particles around particle $a$ (for example, it is on a free surface), then
the interpolation can be compromised. 

The interpolation can be improved by using a Shepard correction \citep{shepard682Dinterp},
originally devised as a low cost improvement to data fitting. This correction
divides the interpolant by the sum of kernel values at the SPH particle positions,
so the summation interpolant becomes

\begin{equation}
A_a = \frac{1}{\sum_b \frac{m_b}{\rho_b} W_{ab}(h_a)} \sum_b m_b \frac{A_b}{\rho_b} W_{ab}(h_a). 
\end{equation}

This correction ensures that a constant field will always be interpolated
exactly close to boundaries, and improves the interpolation accuracy of other, non-constant fields.

\section{SPH-DEM Model}\label{sec:SPHDEMModel}

\subsection{SPH implementation of the AVNS equations}\label{sec:SPH}

SPH is based on a similar local averaging technique as the 
AVNS equations, so it is natural to convert the interpolation integrals in Eqs.\ (\ref{eq:AVNSvariable})
and (\ref{eq:AVNSporosity}) to SPH sums using a smoothing kernel $W(r,h)$ in place of $g(r)$. 

To calculate the porosity $\epsilon_a$ at the center position of SPH/DEM particle $a$, the integral in 
Eq.\ (\ref{eq:AVNSporosity}) is converted into a sum over all DEM particles within the
kernel radius and becomes

\begin{equation}\label{eq:epsilonCalculation}
\epsilon_a = 1 - \sum_{j} W_{aj}(h_c) V_j,
\end{equation}

where $V_j$ is the volume of DEM particle $j$. For readability, sums over SPH
particles use the subscript $b$, while sums over surrounding DEM particles use
the subscript $j$. Note that we have used a coupling smoothing length $h_c$ to
evaluate the porosity, which sets the length scale for the coupling terms
between the phases. Here we set $h_c$ to be equal to the SPH smoothing length,
but in practice this can be set within a range such that $h_c$ is large enough
that the porosity field is smooth but small enough to resolve the important
features of the porosity field. For more details on this point please consult
the numerical results of the test cases and the conclusions of this paper.

Applying the local averaging method to the Navier-Stokes equations, Anderson
and Jackson derived the continuity and momentum equations shown in Eqs.
(\ref{eq:aveContinuity}) and (\ref{eq:aveMomentum2}) respectively. To convert
these to SPH equations, we first define a superficial fluid density $\rho$
equal to the intrinsic fluid density scaled by the local porosity
$\rho=\epsilon \rho_f$.

Substituting the superficial fluid density into the averaged continuity and
momentum equations reduces them to the normal Navier-Stokes equations.
Therefore, our approach is to use the standard weakly compressible SPH
equations, see \citep{robinson11direct}, using the superficial density for the
SPH particle density and adding terms to model the fluid-particle drag.

The rate of change of superficial density is calculated using the variable
smoothing length terms derived by \citet{price12smoothed}.

\begin{equation} \label{Eq:changeInDensity}
\frac{D\rho_a}{Dt} = \frac{1}{\Omega_a}\sum_b m_b \mathbf{u}_{ab} \cdot \nabla_a W_{ab}(h_a),
\end{equation}

where $\mathbf{u}_{ab}=\mathbf{u}_a-\mathbf{u}_b$. The derivative on the lhs of Eq.\ (\ref{Eq:changeInDensity}) denotes the time derivative of the superficial fluid density for each SPH particle $a$. Since the SPH particles move with the flow, this is equivalent to a material derivative of the superficial density $\rho=\epsilon \rho_f$. For more details of the derivation of Eq.\ (\ref{Eq:changeInDensity}) from Eq.\ (\ref{eq:aveContinuity}), the reader is referred to \citet{monaghan05SPH} or \citet{price12smoothed}.  

The correction term $\Omega_a$ is a correction factor due to the gradient of the smoothing length and is given by

\begin{equation}
\Omega_a = 1 - \frac{\partial h_a}{\partial \rho_a} \sum_b m_b \frac{\partial W_{ab}(h_a)}{\partial h_a}.
\end{equation}

Neglecting gravity, the SPH acceleration equation becomes 

\begin{equation}\label{Eq:sphJustPressureForce}
\frac{d\mathbf{u}_a}{dt} = -\sum_b m_b \left [ \left ( \frac{P_a}{\Omega_a \rho_a^2} + \Pi_{ab} \right ) \nabla_a W_{ab}(h_a) + \left ( \frac{P_b}{\Omega_b \rho_b^2} + \Pi_{ab} \right ) \nabla_a W_{ab}(h_b) \right ]  + \mathbf{f}_a/m_a,
\end{equation}

where $\mathbf{f}_a$ is the coupling force on the SPH particle $a$ due to the
DEM particles (see Section \ref{sec:fluidParticleCoupling}). The viscous term
$\Pi_{ab}$ models the divergence of the viscous stress tensor in Eq.\ (\ref{eq:aveMomentum2}) is calculated using the term proposed by \citet{monaghan97SPHRiemannSolvers},
which is based on the dissipative term in shock solutions based on Riemann
solvers. For this viscosity

\begin{equation}\label{Eq:monaghansViscousTerm}
\Pi_{ab} = - \alpha \frac{u_{sig} u_n }{2 \overline{\rho}_{ab} |\mathbf{r}_{ab}|},
\end{equation}

where $u_{sig} = c_s + u_n / |\mathbf{r}_{ab}| $ is a signal velocity that
represents the speed at which information propagates between the particles. The
normal velocity difference between the two particles is given by $u_n =
\mathbf{u}_{ab} \cdot \mathbf{r}_{ab}$. The constant $\alpha$ can be related to
the dynamic viscosity of the fluid $\mu$ using 

\begin{equation}\label{eq:alphaToMu}
\mu = \rho \alpha h c_s / S,
\end{equation}

where $S=112/15$ for two dimensions and $S=10$ for three \citep{monaghan05SPH}.
For some of the reference fluids we have chosen to simulate in this paper it
was found that the physical viscosity was not sufficient to stabilise the
results (see Section \ref{sec:effectOfPorosity}), and it was necessary to add
an artificial viscosity term with $\alpha_{art} = 0.1$. However, this viscosity
term is only applied when the SPH particles are approaching each other (i.e.
$u_{ab}\cdot r_{ab} < 0)$ so that the dissipation due to the artificial
viscosity is reduced while still stabilising the results.

The fluid pressure in Eq.\ (\ref{Eq:sphJustPressureForce}) is calculated using
the weakly compressible equation of state. This equation of state defines a
reference density $\rho_0$ at which the pressure vanishes, which must be scaled
by the local porosity to ensure that the pressure is constant over varying porosity.

\begin{equation}\label{Eq:sphEquationOfState}
P_a = B \left ( \left ( \frac{\rho_a}{\epsilon_a \rho_0} \right )^\gamma - 1 \right ).
\end{equation}

The scaling factor $B$, is free a-priori and is set so that the density
variation from the local reference density is less than 1 percent, ensuring
that the fluid is close to incompressible. For this, in terms of $B$, the
local sound speed is 

\begin{equation}
c_s^2 =  \left. \frac{\partial P}{\partial \overline{\rho}} \right |_{\overline{\rho}=\epsilon_a \rho_0} = \frac{\gamma B}{\epsilon_a \rho_0},
\end{equation}

and the fluctuations in density can be related to the sound speed and velocity of the SPH particles \citep{monaghan05SPH}:

\begin{equation}
\frac{| \delta \rho |}{\rho} = \frac{u^2}{c_s^2}.
\end{equation}

Therefore, in order to keep these fluctuations less than 1\% in a flow where
the maximum velocity is $u_m$ and the maximum porosity is as always
$\epsilon_m=1$, $B$ is set to 

\begin{equation}
B = \frac{100 \rho_0 u_m^2}{\gamma}.
\end{equation}

As the superficial density will vary according to the local porosity, care must
be taken to update the smoothing length for all particles in order to maintain
a sufficient number of neighbour particles. This is referred to as "variable-h"
in this study. The smoothing length $h_a$ is calculated using

\begin{equation}\label{Eq:variableh}
h_a = \sigma \left ( \frac{m_a}{\rho_a} \right )^{1/d},
\end{equation}

where $d$ is the number of dimensions and $\sigma$ determines the resolution of
the summation interpolant. The value used in all the simulation results
presented here is $\sigma = 1.5$. 

Recall that the SPH density is given by $\rho = \epsilon \rho_f$. Assuming
a constant intrinsic fluid density $\rho_f$, the smoothing length $h$
is thus proportional to the local porosity $h \propto (1/\epsilon)^{1/d}$. 

Setting $\epsilon=1$ gives the minimum smoothing length possible in the
simulation. One of the key assumptions of the SPH-DEM method is that the
smoothing length scale $h$ is sufficiently larger than the solid particle
diameter, and the results for the Single Particle Sedimentation tests case
(Section \ref{sec:SPSresolution}) indicate that the minimum $h$ should always be
greater than two times the solid particle diameter (or much smaller, which is
not considered here).


 





\subsection{Discrete Element Model (DEM)}\label{sec:DEM}

In DEM, Newton's equations of motion are
integrated for each individual solid particle. Interactions between the
particles involve explicit force expressions that are used whenever two particles
come into contact.

Given a DEM particle $i$ with position $\mathbf{r}_i$, the equation of motion is 

\begin{equation}
   m_i \frac{d^2 \mathbf{r}_i}{dt^2} = \sum_j \mathbf{c}_{ij} + \mathbf{f}_i +  m_i\mathbf{g},
\end{equation}

where $m_i$ is the mass of particle $i$, $\mathbf{c}_{ij}$ is the contact force
between particles $i$ and $j$ (acting from $j$ to $i$) and $\mathbf{f}_i$ is the fluid-particle coupling
force on particle $i$. For the simulations presented below, we have used the linear spring dashpot
contact model

\begin{equation}
\mathbf{c}_{ij} = -(k \delta -\beta \dot{\delta})\mathbf{n}_{ij},
\end{equation}

where $\delta$ is the overlap between the two particles (positive when the
particles are overlapping, zero when they are not) and $\mathbf{n}_{ij}$ is the
unit normal vector pointing from $j$ to $i$.. The simulation timestep is
calculated based on a typical contact duration $t_c$ and is
given by $\Delta t = \frac{1}{50}t_c$, with $t_c=\pi/\sqrt{(2k/m_i)-\beta/m_i}$.

The timestep for the SPH method is set by a CFL condition

\begin{equation}\label{Eq:CFL}
\delta t_1 \le \min_a \left ( 0.6 \frac{h_a}{u_{sig}} \right ),
\end{equation}

where the minimum is taken over all the particles. This is normally much larger
than the DEM contact time, so the DEM timestep usually sets the minimum
timestep for the SPH-DEM method.

See Table \ref{Tab:parameters} in Section \ref{sec:ValidationTestCases} for all
the parameters and time-scales used in these simulations.

\subsection{Fluid-Particle Coupling Forces}\label{sec:fluidParticleCoupling}

The force on each solid particle by the fluid is \citep{anderson67fluid}

\begin{equation}\label{Eq:demCouplingForce}
\mathbf{f}_i = V_i (-\nabla P + \nabla \cdot \mathbf{\tau})_i + \mathbf{f}_d(\epsilon_i,\mathbf{u}_s),
\end{equation}

where $V_i$ is the volume of particle $i$. The first two terms models the
effect of the resolved fluid forces (buoyancy and shear-stress) on the
particle. For a fluid in hydrostatic equilibrium, the pressure gradient will
reduce to the buoyancy force on the particle. The divergence of the shear
stress is included for completeness and ensures
that the movement of a neutrally buoyant particle will follow the fluid
streamlines. For the simulations considered in this paper this term will not be
significant. 

The force $\mathbf{f}_d$ is a particle drag force that depends on the local
porosity $\epsilon_i$ and the superficial velocity $\mathbf{u}_s$ (defined in
the following section).  This force models the drag effects of the unresolved
fluctuations in the fluid variables and is normally defined using both
theoretical arguments and fits to experimental data. For a single particle in
3D creeping flow this term would be the standard Stokes drag force.  For higher
Reynolds numbers and multiple particle interactions this term is determined
using fits to numerical or experimental data \citep{hoef05lattice}. See Section
\ref{sec:DragLaws} for further details.

The pressure gradient and the divergence of the stress tensor are evaluated at
each solid particle using a Shepard corrected \citep{shepard682Dinterp} SPH
interpolation. Using the already given SPH acceleration equation, Eq.
(\ref{Eq:sphJustPressureForce}), this becomes

\begin{align}
&(-\nabla P + \nabla \cdot \mathbf{\tau})_i = \frac{1}{\sum_b \frac{m_b}{\rho_b} W_{ab}(h_b)} \sum_b m_b \theta_b  W_{ib}(h_b), \\
&\theta_a = -\sum_b m_b \left [ \left ( \frac{P_a}{\Omega_a \rho_a^2} + \Pi_{ab} \right ) \nabla_a W_{ab}(h_a) + \left ( \frac{P_b}{\Omega_b \rho_b^2} + \Pi_{ab} \right ) \nabla_a W_{ab}(h_b) \right ].
\end{align}

In order to satisfy Newtons third law (i.e. the action = reaction principle),
the fluid-particle coupling force on the fluid must be equal and opposite to
the force on the solid particles. Each DEM particle is contained within
multiple SPH interaction radii, so care must be taken to ensure that the two
coupling forces are balanced. 

The coupling force on SPH particle $a$ is determined by a weighted average of
the fluid-particle coupling force on the surrounding DEM particles. The
contribution of each DEM particle to this average is scaled by the value of the
SPH kernel.

\begin{equation}\label{Eq:SPHCoupleForce}
\mathbf{f}_a = - \frac{m_a}{\rho_a} \sum_j \frac{1}{S_j} \mathbf{f}_j W_{aj}(h_c),
\end{equation}

where $\mathbf{f}_j$ is the coupling force calculated for each DEM particle
using Eq.\ (\ref{Eq:demCouplingForce}). The scaling factor $S_j$ is added to
ensure that the force on the fluid phase exactly balances the force on the
solid particles. It is given by

\begin{equation}\label{Eq:SPHCoupleForce2}
S_j = \sum_b{\frac{m_b}{\rho_b} W_{jb}(h_c)},
\end{equation}

where the sum is taken over all the SPH particles surrounding DEM particle $j$.
For a DEM particle immersed in the fluid this will be close to unity. 

\subsection{Fluid-Particle Drag Laws}\label{sec:DragLaws}

The drag force $\mathbf{f}_d$ depends on the superficial velocity
$\mathbf{u}_s$, which is proportional to the relative velocity between the
phases. If $\mathbf{u}_f$ and $\mathbf{u}_i$ are the fluid and particle
velocity respectively, then the superficial velocity is defined as

\begin{equation}
\mathbf{u}_s = \epsilon_i (\mathbf{u}_f-\mathbf{u}_i). 
\end{equation}

This term is used as
the dependent variable in many drag laws as it is easily measured from
experiment by dividing the fluid flow rate by the cross-sectional area.

In the SPH-DEM model, the fluid velocity $\mathbf{u}_f$ used to calculate the
superficial velocity, is found at each DEM particle position using a Shepard
corrected SPH interpolation. The value of the porosity field at each DEM particle position $\epsilon_i$ is found in an identical way.

The simplest drag law is the Stokes drag force

\begin{equation}\label{eq:stokesDrag}
\mathbf{f}_d = 3 \pi \mu d \mathbf{u}_s,
\end{equation}

where $d$ is the particle diameter. This is valid for a single particle in creeping
flow.

\citet{coulson93chemical} proposed a drag law valid for
a single particle falling under the full range of particle Reynolds Numbers
$Re_p  = \rho_f |\mathbf{u}_s| d / \mu$.

\begin{equation}\label{eq:coulson_and_richardson}
\mathbf{f}_d = \frac{\pi}{4} d^2 \rho_f |\mathbf{u}_s| \left (1.84
Re_p^{-0.31}+0.293 Re_p^{0.06} \right )^{3.45}
\end{equation}

For higher Reynolds numbers and multiple particles, the drag law can be generalised to

\begin{equation}\label{eq:singleParticleInInfiniteDomain}
\mathbf{f}_d = \frac{1}{8} C_d f(\epsilon_i) \pi d^2 \rho_f |\mathbf{u}_s|\mathbf{u}_s,
\end{equation}

where $C_d$ is a drag coefficient that varies with the particle Reynolds number
$Re_p = \rho_f |\mathbf{u}_s| d / \mu$, and $f(\epsilon_i)$ is the voidage function
that models the interactions between multiple particles and the fluid.

A popular definition for the drag coefficient was proposed by
\citet{dallavalle48micromeritics}

\begin{equation}\label{eq:DallavalleDrag}
C_d = \left [ 0.63 + \frac{4.8}{\sqrt{Re_p}} \right ]^2.
\end{equation}

Di Felice proposed a voidage function based on experimental data of fluid flow through packed spheres \citep{difelice94voidage}

\begin{align}\label{eq:DiFeliceDrag}
&f(\epsilon_i) = \epsilon_i^{-\xi}, \\
&\xi = 3.7 - 0.65 \exp \left [ -\frac{(1.5 - \log_{10}Re_p)^2}{2} \right
].\label{eq:DiFeliceDrag2}
\end{align}

Both the Stokes drag term (as the simplest reference case) and the combination of
Dallavalle and Di Felice's drag terms are used in the simulations presented in
this paper. Another commonly used drag term is given by a combination of drag
terms by \citet{ergun52fluid} and \citet{wen66mechanics}. For
$\epsilon_i \rightarrow 1$ this term and Di Felice's are identical (over all $Re$). As the
porosity decreases both drag terms generally follow the same trend, although the Ergun
and Wen \& Yu model gives a larger drag force for dense systems.

\section{Validation Test Cases} \label{sec:ValidationTestCases}

In this section, three different sedimentation test cases are proposed and used
to verify that SPH-DEM correctly models the dynamics of the two phases (fluid
and solid particles) and their interactions.

\begin{enumerate}
\item Single Particle Sedimentation (SPS)
\item Sedimentation of a constant porosity block (CPB)
\item Rayleigh Taylor Instability (RTI)
\end{enumerate}

\begin{figure}
\centering
\includegraphics[width=0.7\textwidth]{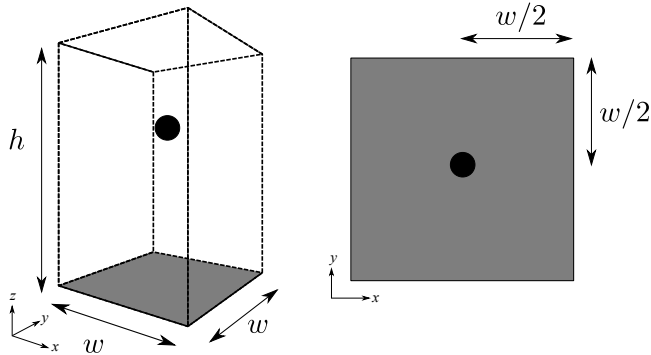}
\caption{Setup for test case SPS, single particle sedimentation in a fluid column. 
(Left) Perspective view, showing the fluid domain, the no-slip bottom boundary and the single spherical DEM particle. 
(Right) Top view, the grey area is the bottom no-slip boundary}
\label{fig:singleDiagram}
\end{figure}

These test cases were designed to test the particle-fluid coupling mechanics in
order of increasing complexity. The first test case simply requires the correct
calculation and integration of the drag force on the single particle, the
single particle being too small to noticeably alter the surrounding fluid
velocity. The second requires that the drag on both phases and the displacement
of fluid by the particles be correctly modelled for a simple velocity field and
constant porosity. The third test case does the same but with a more
complicated and time-varying velocity and porosity field due to the moving
particle phase.

The first test case (SPS) models a single particle sedimenting in a fluid
column under gravity. Figure \ref{fig:singleDiagram} shows a diagram of the
simulation domain. The water column has a height of $h=0.006 \mathrm{m}$ and the bottom boundary is constructed using
Lennard-Jones repulsive particles (these particles are identical to those used
by \citet{monaghan03fluid}). The boundaries in the $x$ and
$y$ directions are periodic with a width of $w=0.004\text{ m}$ and gravity acts in the
negative $z$ direction. The single DEM particle is initialised at $z=0.8h$. It
has a diameter equal to $d = 1\times 10^{-4} \text{ m}$ and has a density $\rho_p = 2500\text{ kg/m}^3$.

For the initial conditions of the simulation, the position of the DEM particle is
fixed and the fluid is allowed to reach hydrostatic equilibrium. The particle
is then released at $t=0\text{ s}$.

\begin{figure}
\centering
\includegraphics[width=0.37\textwidth]{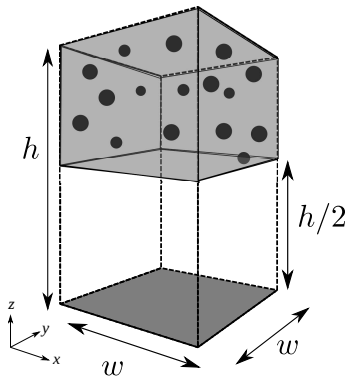}
\caption{Setup for test cases CPB and RTI, multiple particle sedimentation in a
fluid column.}
\label{fig:multipleDiagram}
\end{figure}

Most fluid-particle systems of interest will involve large numbers of
particles, and therefore the second test case (CPB) involves the sedimentation
of multiple particles through a water column. In this case, a layer of
sedimenting particles is placed above a clear fluid region. Figure
\ref{fig:multipleDiagram} shows the setup geometry. The fluid column is
identical to the previous test case, but now the upper half of the column is
occupied by regularly distributed DEM particles on a cubic lattice, with
a given porosity $\epsilon$. The separation between adjacent DEM particles on
the lattice is given by $\Delta r = (V/(1-\epsilon))^{1/3}$, where $V$ is the
(constant) particle volume. The diameter and density of the particles are 
identical to the single particle case. In order to maintain a constant
porosity as the layer of particles falls, the DEM particles are restricted from
moving relative to each other and the layer of particles falls as a block (only
translation, no rotation of the layer).

The third test case (RTI) uses the same simulation domain and initial
conditions as CPB, but now the particles are allowed to move freely. This setup
is similar in nature to the classical Rayleigh-Taylor (RT) instability, where a
dense fluid is accelerated (normally via gravity) into a less dense fluid. The
combination of particles and fluid can be modeled as a two-fluid system with
the upper ``fluid" having an effective density $\rho_d$, and an effective
viscosity $\mu_d$, both higher than the properties of the fluid without
particles. From this an expected growth rate can be calculated for the
instability and compared with the simulated growth rate. See Section
\ref{sec:RTI} for more details. 

For all three test cases, three different model fluids are used to evaluate the
SPH-DEM model at different fluid viscosities and particle Reynolds numbers.
The densities and viscosities of these fluids correspond to the physical
properties of air, water and a 10\% glycerol-water solution.

\subsection{Simulation Parameters, Analytical Solutions and Timescales}

\begin{sidewaystable*}
\caption{Relevant parameters and timescales for the simulations using different fluids. Parameters appearing only in one column are kept constant for all fluids.}
\renewcommand{\arraystretch}{1.3}

\label{Tab:parameters}
\footnotesize{
\begin{center}
\begin{tabular}{|c|c|c|c|c|c|}
\hline
\bf{} & \bf{Notation} & \bf{Units} & \bf{Air} & \bf{Water}& \bf{Water + 10\% Glycerol}\\
\hline
Box Width& $w$ &$\text{m}$& $4 \times 10^{-3}$ &&\\
Box Height& $h$ &$\text{m}$& $6 \times 10^{-3}$ &&\\
\hline
Fluid Density & $\rho$ &$\text{kg/m}^3$& $1.1839$ & $1000$ & $1150$ \\
Fluid Viscosity & $\mu$ &$ \text{Pa} \cdot \text{s}$&$1.86 \times 10^{-5}$ & $8.9 \times 10^{-4}$ & $8.9\times 10^{-3} $  \\
\hline
Particle Density & $\rho_p$ &$ \text{kg/m}^3$&$2500$ &&\\
Particle Diameter & $d$ &$\text{m}$& $1.0 \times 10^{-4}$ &&\\
Spring Stiffness& $k$ &$ \text{kg/s}^2$& $1.0 \times 10^{-4}$ &&\\
Spring Damping& $\beta$ &$\text{kg/s}$& $0 $ &&\\
\hline
Porosity & $\epsilon$ && 0.6-1.0 &&\\
Calculated Terminal Velocity (Eq.\ \ref{eq:expectedDiFeliceTermVel})& $|\mathbf{u}_t|$& $\text{m/s}$&0.102-0.5 &$1.3\times 10^{-3}$-$7.6\times 10^{-3}$ &$ 1.3\times 10^{-4}$-$8.4\times 10^{-4}$ \\
Calculated Terminal Re Number (Eq.\ \ref{eq:expectedDiFeliceTermVel})& $Re_p$ && 0.65-3.19 & 0.15-0.85 & 0.002-0.011 \\
Archimedes Number (single particle)& $Ar$ && 83.89 &18.57& 0.192 \\ 
\hline
Particle Contact Duration& $t_c$ &$\text{s}$& $2.54\times 10^{-3}$ && \\
Fluid CFL Condition& $t_f$ &$\text{s}$& 1.4-4.5 $\times 10^{-5}$ && \\
Fluid-particle Relaxation Time& $t_d$ &$\text{s}$& $7.47 \times 10^{-2} $ & $1.56 \times 10^{-3}$ & $1.56 \times 10^{-4}$ \\
\hline
\end{tabular}
\end{center}
}
\end{sidewaystable*}

Table \ref{Tab:parameters} shows the parameters used in the three test cases.
Each column corresponds to a different model fluid. Where a value appears only
in one column, this indicates that the parameter is constant for all the
fluids. The particle Reynolds number is calculated using the expected terminal
velocity of either the single particle or porous block. 

The standard Stokes law, Eq.\ (\ref{eq:stokesDrag}), can be used to calculate the
vertical speed of a single particle falling in a quiescent fluid. 

\begin{equation}\label{Eq:fallingParticleVel}
v(t) = \frac{(\rho_p-\rho) V g}{b} \left ( 1-e^{-bt/m} \right ), \textrm{  with constant } b =  3 \pi \mu d.
\end{equation}

Since we are interested in a range of particle Reynolds numbers, not just at
the Stokes limit, we also consider the Di Felice drag force, Eq.\
(\ref{eq:DiFeliceDrag}), which is valid for higher Reynolds numbers and varying
porosity (i.e.\ it considers the interaction of multiple particles). When the buoyancy and gravity force on the falling particle balance out the drag force,
the particle is falling at its terminal velocity. Equating these terms leads to a
polynomial equation in terms of the particle Reynolds number at terminal
velocity

\begin{equation}\label{eq:expectedDiFeliceTermVel}
0.392Re_p^2 + 6.048Re_p^{1.5} + 23.04Re_p - \frac{4}{3}Ar\epsilon^{1+\xi} = 0,
\end{equation}

where $\xi$ is given in Eq.\ (\ref{eq:DiFeliceDrag2}) and $Ar = d^3\rho(\rho_p -
\rho)g/\mu^2$ is the Archimedes number. The Archimedes number gives the ratio
of gravitational forces to viscous forces. A high $Ar$ means that the system is
dominated by convective flows generated by density differences between the
fluid and solid particles. A low $Ar$ means that viscous forces dominate and the
system is governed by external forces only.

Solving for $Re_p$, one can find the expected terminal velocity using $Re_p = \rho |\mathbf{u}_t| d / \mu$.

Note that a range of porosities is used for test cases CPB and RTI, and this
results in a range of particle Reynolds numbers as the terminal velocity
depends on the porosity. 

Also included in Table \ref{Tab:parameters} are the relevant timescales for the
simulations. The particle contact duration $t_c$ and fluid CFL condition $t_f$ are
described in Sections \ref{sec:DEM} and \ref{sec:SPH} respectively. The
fluid-particle relaxation time is the characteristic time during which a
falling particle in Stokes flow will approach its terminal velocity. This is
given by $t_d=m/b$ from Eq.\ (\ref{Eq:fallingParticleVel}). This relaxation time provides
another minimum timestep for the SPH-DEM simulation, given by

\begin{equation}\label{eq:relaxationCondition}
\Delta t_{relax} \le \frac{1}{20} \frac{m}{b}.
\end{equation}

The physical properties of the solid DEM particles are constant over all the
simulated cases. Since the results of the test cases are insensitive to the
particle-particle contacts, a relatively low spring stiffness of $k= 10^{-4}
\text{kg/s}^2$ was used. This value ensures that the timestep is limited by the
fluid CFL condition, rather than the DEM timestep, significantly speeding up
the simulations. 

\section{Single Particle Sedimentation (SPS)}\label{sec:SPS}

This section describes the results from SPH-DEM simulations using the first
test case (SPS). We tested one and two-way coupling between the phases, the
effect of different drag laws (Stokes and Di Felice), different fluid
properties (air, water and water-glycerol) and the effect of varying the fluid resolution.

\subsection{One and two-way coupling in Stokes flow}

For a single particle falling in Stokes flow the standard Stokes drag equation,
Eq.\ (\ref{eq:stokesDrag}), can be used. Since Stokes drag law
assumes a quiescent fluid, the force on the fluid due to the particle is set to
zero ($\mathbf{f}_a=0$ in Eq.\ (\ref{Eq:SPHCoupleForce})). This implements a
one-way coupling between the phases. Note that the SPH particles can still
interact with the DEM particles through the porosity field, but for a single
particle this effect will be negligible.

\begin{figure}
\centering
\includegraphics[height=0.9\textwidth, angle=-90]{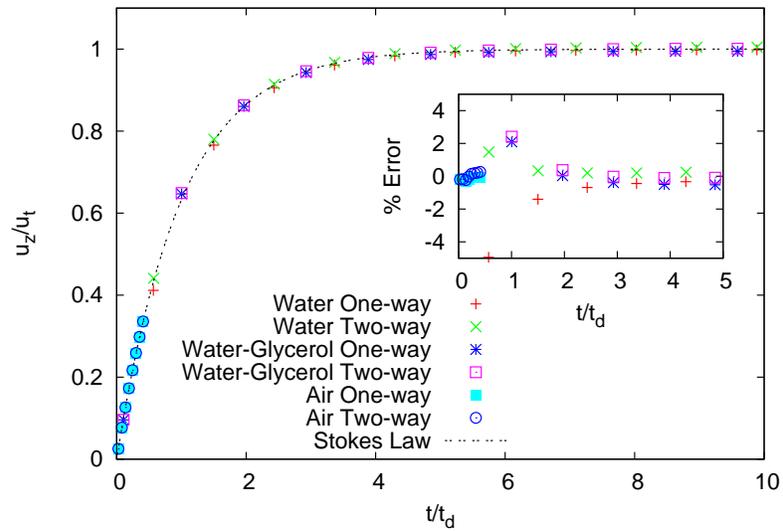}
\caption{Normalised sedimentation velocity as a function of scaled time for a
single particle in different fluids falling from rest with both one-way and two-way coupling. The dashed line is the theoretical result integrating Stokes law.
The particle's vertical velocity is scaled by the expected terminal velocity
$|\mathbf{u}_t|$ and time is scaled by the drag relaxation time $t_d$. The inset
shows the percentage error between the SPH-DEM and the expected trajectory. The
fluid resolution is set to $h=6d$, where $d$ is the particle diameter.}
\label{fig:SPSoneway_water}
\end{figure}

In Figure \ref{fig:SPSoneway_water} the evolution of a DEM particle's
vertical speed in water is shown for one-way and two-way coupling. Also shown is
the expected analytical prediction using Eq.\ (\ref{Eq:fallingParticleVel}). The
falling DEM particle reproduces the analytical velocity very well for both
one-way and two-way coupling and the error between the two curves is less than
1\% for the vast majority of the simulations.  Note that the initial error curve
reaches 5\% when the particle is first released, but this is is a
short-lived effect and the error drops below 1\% after a time of about 
$t_d$, the relaxation time for the drag force.

These results indicate that the pressure gradient, calculated from the SPH
model, very accurately reproduces the buoyancy force on the particle, balancing
out the drag force at the correct terminal velocity.  The results are close
for both one-way and two-way coupling, indicating that the drag force on the
fluid has a negligible effect here. This is true as long as the fluid resolution
is sufficiently larger than the DEM particle diameter (this is explored in
more detail in Section \ref{sec:SPSresolution}).

Figure \ref{fig:SPSoneway_water} also shows the same result for a DEM particle
falling in air and in the water-glycerol mixture. 

For air, the drag force on
the particle is much lower than for water, and the particles do not have time
to reach their terminal velocity before reaching the bottom boundary, where the
simulation ends. As for the previous simulation with water, there is initially
a larger (approx 4\%) underestimation of the particle vertical speed, but once
again this occurs only for a very small time period and does not affect the
long term motion of the particle. For the majority of the simulation the error
is less than 1\% for both one-way and two-way coupling.

The results for the water-glycerol fluid are qualitatively similar to water.
Here the drag force on the particle is much higher than for water and the
particle reaches terminal velocity very quickly.  As long as the
simulation timestep is modified to resolve the drag force relaxation time $t_d$
as per Eq.\ (\ref{eq:relaxationCondition}), the results are accurate. For both
the one-way and two-way coupling, the simulated velocity matches the analytical
velocity very well and the error remains less than 1\% for the duration of
the simulation. 

In summary, the results for the one-way and two-way coupling between the fluid
and particle for all the reference fluids are very accurate, and reproduce the
analytical velocity curve within 1\% error besides short-lived higher
deviations at the initial onset of motion. All data scale using $u_t$ and
$t_d$ for velocity and time, respectively.

\subsection{Effect of Fluid Resolution} \label{sec:SPSresolution}

\begin{figure}
\centering
\includegraphics[height=0.9\textwidth,
angle=-90]{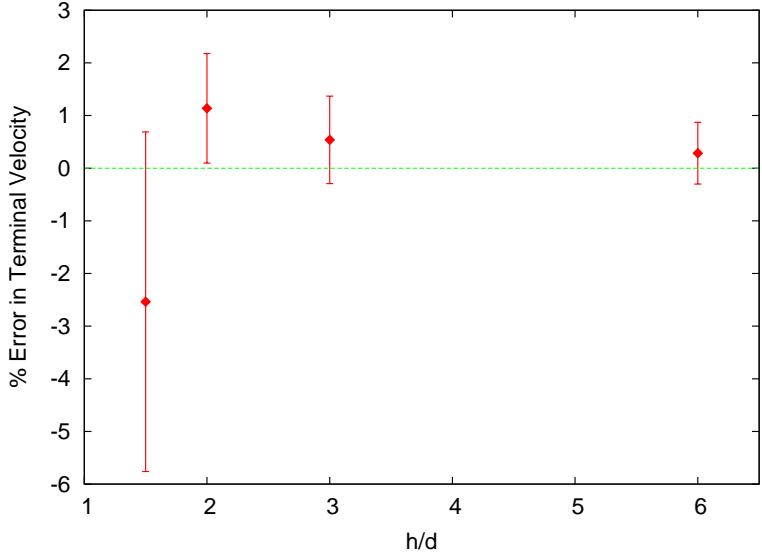}
\caption{The effect of fluid resolution for the SPS test case, with water as the surrounding fluid. 
The average percentage error between the particle terminal velocity and the
analytical value is plotted against $h/d$, where $h$ is the SPH resolution and
$d$ is the DEM particle diameter. The errorbars show one standard deviation from
the mean.}
\label{fig:termVelTwoWay_res}
\end{figure}

In this section we vary the fluid resolution to see its effects on the SPS
results. Using water as the reference fluid, four different simulations were
performed with the number of SPH particles was ranging from 10x10x15 particles
to 40x40x60. Using the SPH smoothing length $h$ as the resolution of the fluid,
this gives a range of $1.5d \le h \le 6d$, where $d$ is the DEM particle
diameter. 

Figure \ref{fig:termVelTwoWay_res} shows the percentage difference between the
average terminal velocity of the particle and the expected Stokes law. The
error bars in this plot show one standard deviation of the fluctuations in the
terminal velocity around the average, taken over a time period of $0.34$ s after the terminal velocity has
been reached.

The $h/d=6$ resolution corresponds to that used in the previous one- and two-way
coupled simulations, and the percentage error here is similar to the one-way
case, which is a mean of 0.2\% with a standard deviation of 0.8\%. As the fluid
resolution is increased there is no clear trend in the average terminal
velocity, but there is an obvious increase in the fluctuation of the terminal
velocity around this mean. For $h/d \ge 2$, the standard deviation of these
fluctuations is less than 1\%, but this quickly grows to 3\% for $h/d=1.5$. 

The increased error as the fluid resolution approaches the particle diameter is
due to one of the main assumptions of the AVNS equations, i.e. that the fluid
resolution length scale is sufficiently larger than the solid particle diameter.
In this case the smoothing operator used to calculate the porosity field is also
much greater than the particle diameter and this will result in a smooth
porosity field. As the fluid resolution is reduced to the particle diameter the
calculated porosity field will become less smooth and there will emerge local
regions of high porosity at the locations of the DEM particles.
Therefore, the fluctuations in the porosity field become greater which will
cause greater fluctuations in the forces on the SPH particles leading to a more
noisy velocity field.

Another trend (not clear in Figure \ref{fig:termVelTwoWay_res} but can be seen
for higher density solid particles) is the terminal velocity of the particle
increasing with increasingly finer fluid resolution.  Due to the two-way
coupling, the drag force on the particle will be felt by the fluid as an equal
and opposite force. This will accelerate the fluid particles by an amount
proportional to the relative mass of the SPH and DEM particles. For higher resolutions the
mass of the SPH particles is lower, leading to an increase in vertical velocity
of the affected fluid particles. Since the DEM particle's drag force depends on
the velocity difference between the phases, which is now smaller, this will lead
to a increase in the particle's terminal velocity. For the SPS test case shown
here, the single particle does not exert too much force on the fluid and this is
not a very large effect. As the fluid resolution is increased from $h/d=6$ to
$2$, there is a slight increase (on the order of 1-2\%) in the terminal
velocity; for lower $h/d$ the trend is lost, likely due to
the increasing noise due to the fluctuations in the porosity field.

\subsection{The effect of fluid properties and particle Reynolds number}

We have used three different reference fluids in the simulations, corresponding
to air, water and a water-glycerol mixture. Using the SPS test case, this
results in a range of particle Reynolds numbers between $0.011$ (water-glycerol)
and $3.19$ (air), allowing us to explore a realistic range of particle Reynolds
numbers. We have further extended this range by considering two additional
(artificial) fluids with a density of water but lower viscosities, resulting in
a range of $0.011 \le Re_p \le 9$.

Rather than assuming Stokes flow as in the previous sections, here we will use
the Di Felice drag law ($\epsilon=1$), which is assumed to be valid for
all Reynolds numbers. This will be compared against fully resolved simulations
using COMSOL Multiphysics (finite element analysis, solver and simulation
software. \url{http://www.comsol.com/}).

\begin{figure}
\centering
\includegraphics[height=0.9\textwidth,angle=-90]{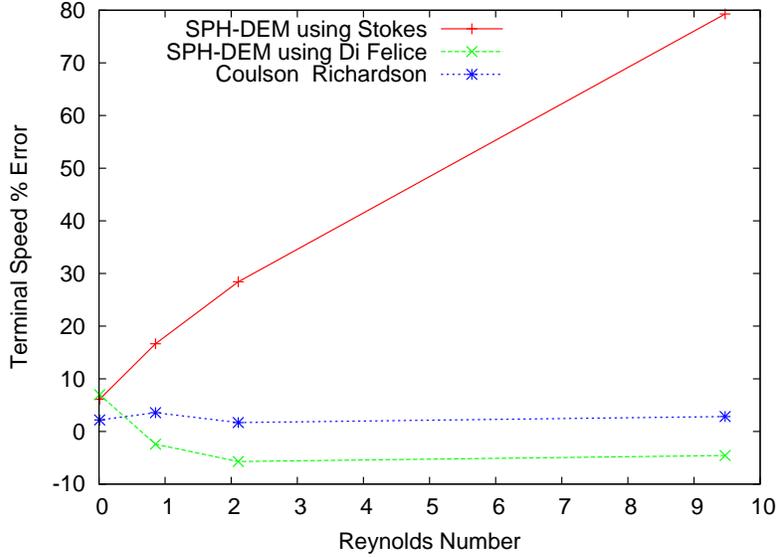}
\caption{Error in SPH-DEM average SPS terminal velocity at different terminal Re
numbers. The fully resolved COMSOL simulation is used as reference for the error
calculation. The solid red and dashed green lines show the results using
either Stokes or Di Felice drag law. The dotted blue line shows the
reference terminal velocity calculated using the Coulson and Richardson drag law
\citep{coulson93chemical}. The SPH-DEM results use a fluid resolution of
$h/d=6$}
\label{fig:diFeliceCompare}
\end{figure}

Figure \ref{fig:diFeliceCompare} shows the average error in the terminal
velocity measured from the SPH-DEM simulations using both the Stokes and Di
Felice drag laws, using the COMSOL results as the reference terminal velocity.
Since the two drag laws are equivalent at low $Re_p$, they give the same result
at $Re_p = 0.01$. As $Re_p$ increases, the plots diverge, and the simulated
terminal velocity using the Stokes drag quickly becomes much larger than the COMSOL
prediction (as expected since the Stokes drag law is only valid for low $Re_p$).
In contrast, the Di Felice drag law results in a simulated terminal velocity that
follows the same trend as the COMSOL results. At low $Re_p$ the DEM particle
falls slightly ($\sim 5$\%) faster, at higher $Re_p$ it falls slightly (3-6\%)
slower.

For further comparison, the COMSOL results have also been compared with the
analytical drag force model proposed in \cite{khan1987resistance,coulson93chemical} and
reproduced in Eq.\ (\ref{eq:coulson_and_richardson}). The expected terminal velocity was
calculated using this model and plotted alongside the SPH-DEM results in Figure
\ref{fig:diFeliceCompare}. As shown, the COMSOL results agree with this
analytical terminal velocity to within 3.5\% over the range of $Re_p$
considered.

While the results in previous SPS sections have shown that the SPH-DEM model can
accurately (within 1\%) reproduce the expected terminal velocity assuming a
given drag law (Stokes), this subsection illustrated that the final accuracy is
still largely determined by the suitability of the underlying drag law chosen. However, a
full comparison of the numerous drag laws currently in the literature is beyond
the scope of this paper, and for the purposes of validating the SPH-DEM model
we can assume that the chosen drag law (from here on the Di Felice),
approximates well the true drag on the particles. 

\section{Sedimentation of a Constant Porosity Block (CPB)}\label{sec:CPB}

This section shows the results from the Constant Porosity Block (CPB) test
case. In a similar fashion to the SPS case, we explore the effect of fluid
resolution and fluid properties. In addition, we consider the influence of a
new parameter, the porosity of the block, on the results. All the simulations
in this section use two-way coupling, as the hindered fluid flow due to the presence of
the solid particles is an important component of the simulation. As the
porous block falls, the fluid will be displaced and flow upward through the
block, affecting the terminal velocity. All the simulations use the Di
Felice drag law, which is necessary to incorporate the effects of
moderate Re and of neighbouring particles (lower porosity) on the drag force.

\begin{figure}
\centering
\includegraphics[width=0.9\textwidth]{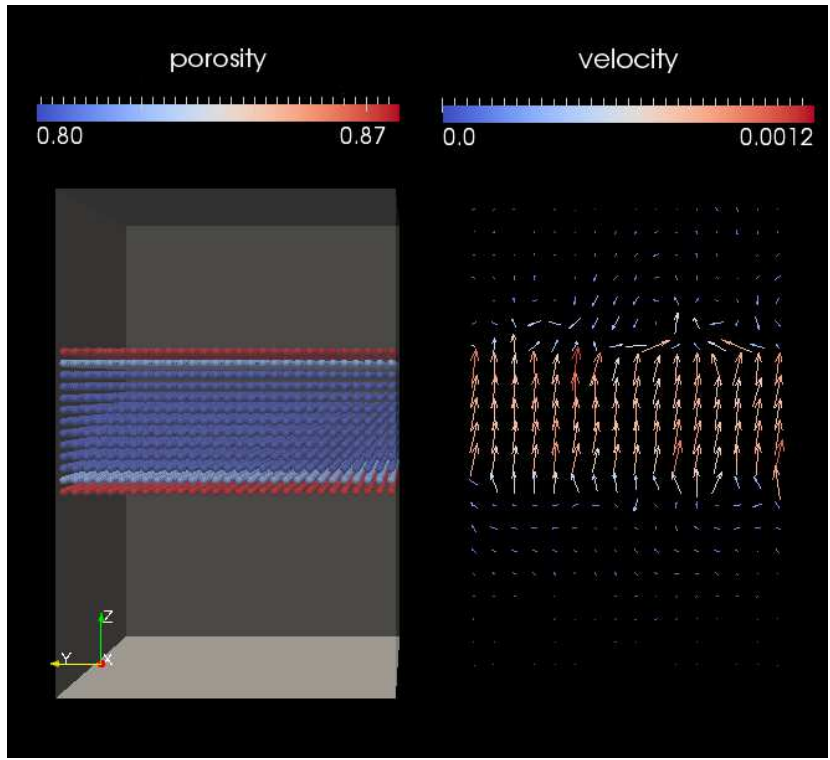}
\caption{Visualisation of the DEM particles for the Constant Porosity Block test
case. On the left the DEM particles are shown coloured by porosity $\epsilon_i$,
and a transparent box representing the simulation domain. On the right the
corresponding fluid velocity field is shown at $x=0$, with the arrows scaled
and coloured by velocity magnitude.}
\label{fig:CPBimage}
\end{figure}

Figure \ref{fig:CPBimage} shows an example visualisation
during the simulation of a block with porosity
$\epsilon=0.8$ falling in water. On the left hand side of the image are shown
the DEM particles (coloured by porosity $\epsilon_i$) falling in the fluid column. The porosity of
most of the DEM particles is $\epsilon=0.8$, as expected, except near the edge
of the block where the discontinuity in particle distribution is smoothed out
by the kernel (with smoothing length $h_c \cong 6d$) in Eq.\ (\ref{eq:epsilonCalculation}). This results in a porosity
greater than 0.8 for DEM particles whose distance is lower than $h_c$ from the edge of
the block. We will show in subsection \ref{sec:CPBresolution} that this effect
can be limited/avoided by choosing a smaller smoothing length.

On the right hand side a vector plot of the velocity field at $x=0$ shows the
upward flow of fluid due to the displacement of fluid by the particles as they
fall. Also noticeable are the fluctuations in velocity near the edges of the
block, which are discussed in more detail in subsection
\ref{sec:effectOfPorosity}.


Shortly after release, the vertical velocity of the CPB converges to a terminal
velocity that is consistent with the expected terminal velocity, although it is
slightly (less than 5\%) higher than expected. The systematically increased
terminal velocity is due to reduced drag at the edges of the block due to the
finite width of the smoothing kernel. As the width of the smoothing kernel $h$
used to calculate the porosity field is larger (by a factor of 2-6, see Figure
\ref{fig:multiResolution_water} for details) than the particle diameter $d$, the
porosity field near the edges of the CPB will be smoothed out according to the
width of the kernel. This results in a slightly higher apparent local porosity
and a reduced drag than what would be expected with $\epsilon=0.8$.

\subsection{The effect of fluid resolution}\label{sec:CPBresolution}

\begin{figure}
\centering
\includegraphics[height=0.9\textwidth,angle=-90]{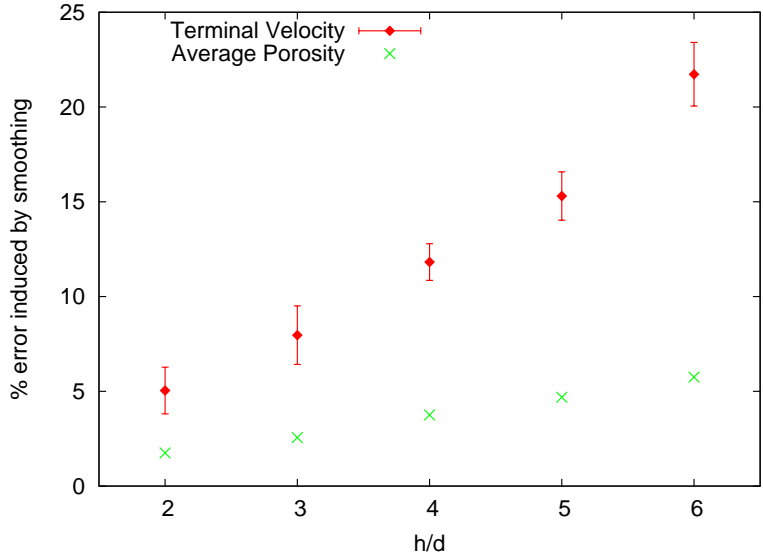}
\caption{Average percentage error in the terminal velocity and average porosity
of the Constant Porosity Block (CPB), with $\epsilon=0.8$ in water, for varying
fluid resolution.
Errorbars in the terminal velocity points show one standard deviation of the vertical velocity data from the
average, taken over a time period of $0.34$ s ($\approx 50 t_d$) after the
terminal velocity has been reached.}
\label{fig:multiResolution_water}
\end{figure}

Figure \ref{fig:multiResolution_water} shows the percentage difference between
the vertical velocity of the block and the expected terminal velocity. The
results from five different simulations are shown, each with a different fluid
resolution ranging from $h/d=6$ to $h/d=2$. The porosity is set to
$\epsilon=0.8$.
The $h/d=6$ simulation suffers from a too strong smoothing of the porosity field
near the edges of the block. Integrating the porosity field over the volume
of the CPB leads to a porosity of 0.85, about 6\% higher than the true porosity of
the block. This results in an increase of 22\% in the terminal velocity of the
block. Increasing the fluid resolution to $h/d=5$ causes the error to decrease
to 15\%, since the interpolated porosity at the edge of the block is now closer
to the set value of $\epsilon=0.8$. Further increases in the fluid resolution
consistently decrease the measured terminal velocity until at $h/d=2$ the error
is only 5\% of the expected value.
These results illustrate how the smoothing applied to the porosity field can
have dramatic results on the accuracy of the simulations. This is largely due to
the fact that the modelled drag only depends on the local (smoothed) porosity,
which does not properly consider sharp porosity gradients.
Thus, the accuracy of the drag law near large changes in porosity is highly
dependent on the magnitude of smoothing applied to the porosity field. This is
true for the Di Felice law and the most other drag laws proposed in the
literature, but there has been some recent work by \citet{xu07discrete}, which
attempts to account for the influence of the porosity gradient, but we will not
study this further here.

\subsection{The effect of porosity}\label{sec:CPBporosity}
\begin{figure}
\centering
\includegraphics[height=0.9\textwidth,angle=-90]{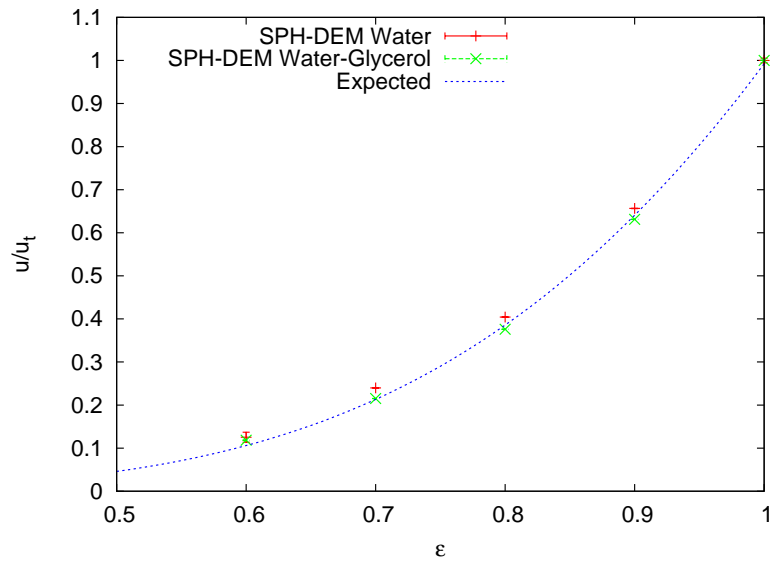}
\caption{Average terminal velocity (scaled by $|\mathbf{u}_t|$, the expected terminal velocity of a single DEM particle) of the Constant
Porosity Block (CPB) in water and water-glycerol for varying porosity and $h/d =
2$. Errorbars show one standard deviation of the vertical velocity data from the
average, taken over a time period of $0.34$ s ($\approx 50 t_d$). The y-axis is 
scaled by $|\mathbf{u}_t|$, the expected terminal velocity of a single DEM
particle given by Eq.\ (\ref{eq:expectedDiFeliceTermVel}), which corresponds to the SPS test case.}
\label{fig:multiPorosity_water}
\end{figure}

Varying the porosity of the CPB allows us to evaluate the accuracy of the
SPH-DEM model at different porosities when $h/d=2$. Figure
\ref{fig:multiPorosity_water} shows the average terminal velocity of the block,
as measured from SPH-DEM simulation of the CPB over a range of porosities from
$\epsilon=0.6$ to $1.0$. Results using both water and water-glycerol as the
interstitial fluid are shown on the same plot by scaling the y-axis by the
expected terminal velocity of a single DEM particle. The average velocity is
taken after the block has reached a steady terminal velocity and the error bars
show one standard deviation of the vertical velocity from the average.

Shown with the SPH-DEM results is the expected terminal velocity computed using
Eq.\ (\ref{eq:expectedDiFeliceTermVel}) and the input porosity of the block. The
SPH-DEM results for both water and water-glycerol match this reference line very
well over the range of porosities tested. At lower porosities the vertical
velocity of the CPB suffers from increasing fluctuation around the mean. This is
a consequence of fluctuations seen in the surrounding fluid velocity, and will
be described further in Section \ref{sec:effectOfPorosity}.

In summary, the simulated terminal velocity for the CPB matched the expected
value over the range of resolutions and porosities considered, as long as the
resolution of the fluid phase (set by $h$) is sufficient to resolve the
porosity field of the given problem. For the CPB we have an discontinuous jump
at the edges of the block from the given porosity of the block to the surrounding
$\epsilon=1$. We found that as long as the fluid resolution was kept at $h=2d$,
where $d$ is the DEM particle diameter (i.e., the length scale of the porosity
jump), the results matched the theoretical predictions within 5\%
over prediction. Using $h < 2d$ is not recommended due to errors caused by 
a non-smooth porosity field, as shown by the SPS test results in Section \ref{sec:SPS}.

\subsection{Effect of Porosity Gradients on Fluid Solution}\label{sec:effectOfPorosity}

\begin{figure}
\centering
\includegraphics[height=0.9\textwidth,angle=-90]{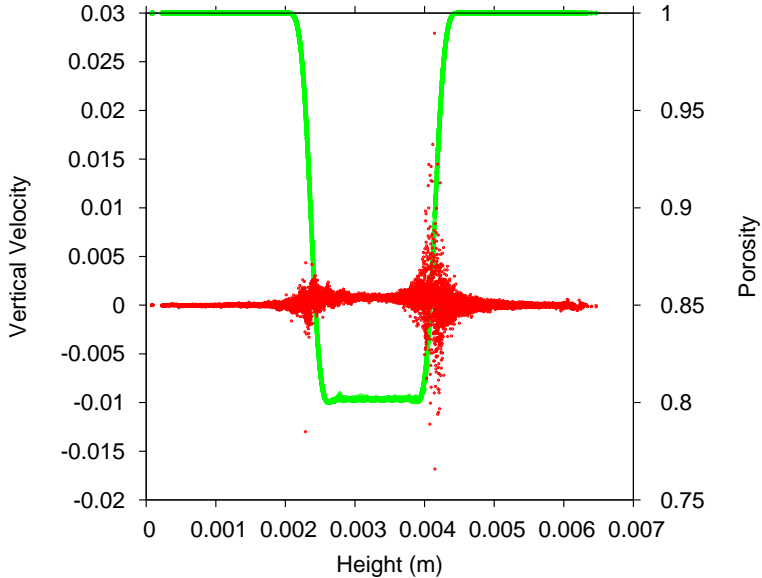}
\caption{Scatter-plot of the vertical velocity (red dots) and porosity (green
line) versus height for all the SPH particles. The test case was CPB with a
porosity of $\epsilon = 0.8$ in water as the surrounding fluid, the fluid
resolution was $h/d = 2$ and $\alpha_{art}=0.1$. Snapshot is taken once the
CPB has reached terminal velocity.}
\label{fig:velAndPor}
\end{figure}

In the previous section it was shown how the smoothing of the porosity
discontinuity of the block slightly affected the drag on the DEM particles and the final
terminal velocity of the block. In this section we will show how the high
porosity gradients near the edge of the block also give rise to further effects
on the SPH solution for the fluid. 

Figure \ref{fig:velAndPor} shows the vertical velocity and porosity for all the
SPH particles in a CPB simulation with fluid resolution $h/d=2$ and
porosity $\epsilon=0.8$, plotted against the vertical position
of the SPH particles. The porosity is rather smooth and clearly shows the location of the CPB. However, there are fluctuations in the
vertical velocity of the SPH particles near the edges of the block, much larger
than the rather small average (positive) velocity inside the block.
These fluctuations are present to different degrees in all of the SPH-DEM simulations
and their magnitude is proportional to the local porosity gradient. Therefore, their
effect is strongest for the simulations with low porosity or fine
fluid resolution (i.e. small $h$).

Given the correlation of these fluctuations with high porosity gradients, their
source is likely to be due to errors in the SPH pressure field. It is
well-known, e.g. \citep{colagrossi03numerical}, that SPH solutions can exhibit
spurious fluctuations in the pressure field, which normally have little or
no effect on the fluid velocity. For our simulations the pressure
of each SPH particle is proportional to $(\rho/\epsilon \rho_0)^7$ and is therefore
very sensitive to changes in $\epsilon$. It is likely that for high
porosity gradients the pressure variations that are normally present would be amplified and
generate corresponding large fluctuations in the velocity field.

As long as the fluctuations do not grow too large, they do not affect the mean
flow of the fluid, as evidenced by the reproduction of the expected terminal
velocity in the previous sections. To ensure the simulation accuracy, it was
found that the application of an artificial viscosity with strength
$\alpha_{art} = 0.1$, see Eq.\ (\ref{Eq:monaghansViscousTerm}), was enough to
damp out the fluctuations in velocity so that they did not have a significant
effect on the results. This value of $\alpha_{art}$ was used in all of the CPB
simulations shown here. The artificial viscosity has little effect on the
settling velocity of the SPS or CPB since this viscosity is only applied between
SPH particles and is not included in the fluid-particle coupling term (Eq.
\ref{Eq:demCouplingForce}). However, for systems where the fluid viscosity plays
an important role (e.g. the Rayleigh Taylor instability), this has an effect
which will be described in the next section.

\section{Rayleigh-Taylor Instability (RTI)} \label{sec:RTI}

\begin{figure}
\centering
\includegraphics[height=0.6\textwidth]{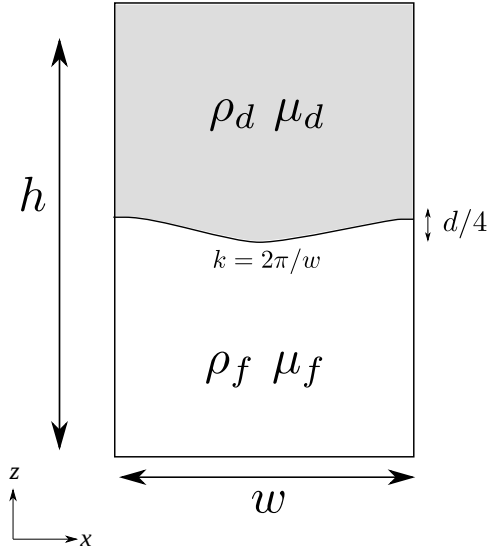}
\caption{Diagram showing a cross-section of the initial setup for the
Rayleigh-Taylor Instability (RTI) test case. The upper grey area is the
particle-fluid suspension with effective density and viscosity $\rho_d$ and
$\nu_d$, the lower white region is clear fluid with density and viscosity
$\rho_f$ and $\nu_f$. The suspension is given an initial vertical perturbation
with wave number $k$ and amplitude $d/4$.}
\label{fig:rayleightaylor_diagram}
\end{figure}

The classic Rayleigh-Taylor fluid instability is seen when a dense fluid is
accelerated into a less dense fluid, for example, under the action of gravity.
Consider a water column of height $h$ filled with a dense fluid with density
$\rho_d$ and viscosity $\nu_d$ located above a lighter fluid with parameters
$\rho_f$ and $\nu_f$. For the RTI test case, the lower and higher density
fluids are represented by the pure fluid and the suspension, respectively. If
the height of the interface between the two fluids is perturbed by a normal mode
disturbance with a certain wave number $k$ (see Figure
\ref{fig:rayleightaylor_diagram} and Eq.\ (\ref{eq:normal_mode_disterbance})),
then this disturbance will grow exponentially with time.

The two-fluid model of a Rayleigh-Taylor instability was derived in the
authoritative text by \citet{chandrasekhar61hydrodynamic}. The
exponential growth rate $n(k)$ of a normal mode disturbance with wave number
$k$ at the interface between the two fluids (with zero surface tension) is
characterised by the dispersion relation \citep{chandrasekhar61hydrodynamic}
given by

\begin{align}
&- \left [ \frac{gk}{n^2} (\alpha_f - \alpha_d) + 1 \right ] (\alpha_c q_d + \alpha_f q_c - k) - 4k \alpha_f \alpha_d \nonumber \\
&+ \frac{4k^2}{n} (\alpha_f \nu_f - \alpha_d \nu_d) [\alpha_d q_f - \alpha_f q_d + k(\alpha_f - \alpha_d)] \nonumber \\
&+ \frac{4k^3}{n^2} (\alpha_f \nu_f - \alpha_d \nu_d)^2 (q_f - k)(q_d-k) = 0,
\label{eq:dispersionRT}
\end{align}

where $\nu_{f,d}=\mu_{f,d}/\rho_{f,d}$ is the kinematic viscosity of the two
phases, $\alpha_{f,d} = \rho_{f,d}/(\rho_f+\rho_d)$ is a density factor and
$q^2_{f,d}=k^2+n/\nu_{f,d}$ is a convenient abbreviation.

For this test case, we use an identical initial condition as in the CPB test
case, with a block of particles immersed in the fluid with an initial
porosity of $\epsilon=0.8$. Using the density of the surrounding fluid $\rho_f$,
the effective density of the fluid-particle suspension is $\rho_d=\epsilon
\rho_f + (1-\epsilon) \rho_p$.

The effective viscosity of the suspension $\mu_{d}$ is estimated here using
Krieger's hard sphere model \citep{krieger59mechanism} (assumed to be valid for
both dilute and dense suspensions)

\begin{equation}\label{Eq:krieger}
\mu_{d} = \mu_{f} \left ( \frac{\epsilon-\epsilon_{min}}{1-\epsilon_{min}}
\right )^{-2.5(1-\epsilon_{min})},
\end{equation}

where $\epsilon_{min} = 0.37$ is the porosity at the maximum packing of the solid particles. 

We generate an initial disturbance in the interface between the two ``fluids" by
adding a small perturbation to the vertical position of every DEM particle

\begin{equation}\label{eq:normal_mode_disterbance}
\Delta z_i = -\frac{d}{4} (1-\cos(k_x x_i))(1-\cos(k_y y_i)),
\end{equation}

where $k_x=k_y=2\pi/w$ and $x_i$ and $y_i$ are the coordinates of particle $i$.
This yields a symmetric disturbance in the interface with a wave length equal to
the box width $w$ and identical to the wave length of the dominant mode.

\begin{figure}
\centering
\includegraphics[width=0.9\textwidth]{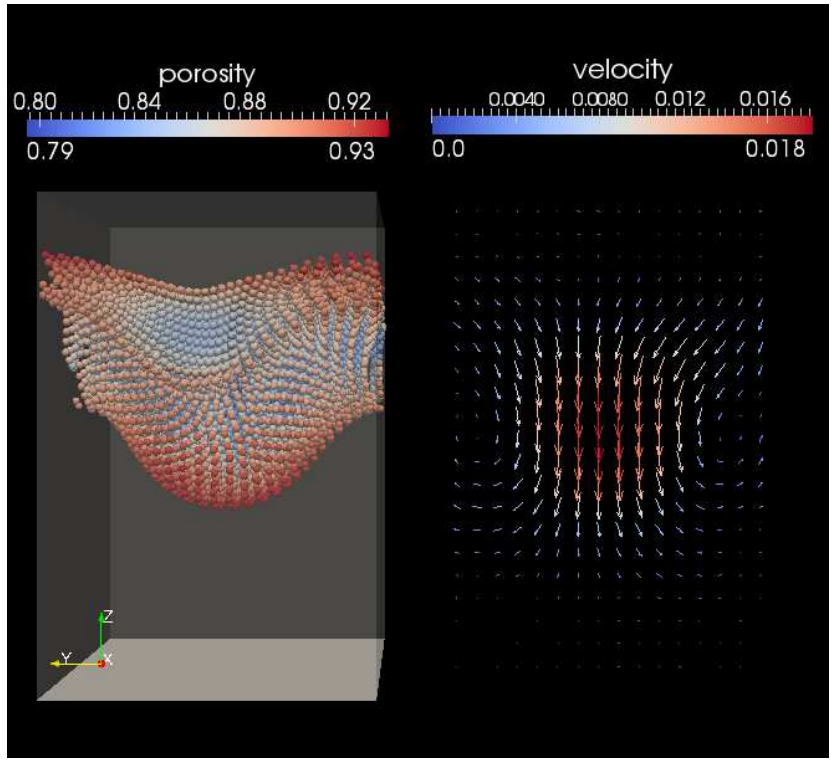}
\caption{Visualisation of the DEM particles (left) and the fluid velocity field
(right) at $x=0$ in the y-z plane, for the Rayleigh Taylor (RT) test case at
$t=0.37$, using $\epsilon=0.8$ and water-glycerol as the surrounding fluid.
The growth rate for this simulations versus time can be seen in Figure \ref{fig:multiGrowthRate_water_glycerol}.}
\label{fig:RTVis}
\end{figure}

Figure \ref{fig:RTVis} shows the positions of the DEM particles during the
growth of the instability, along with the fluid velocity field at $x=0$.
At this time there is a strong fluid circulation that is moving downward in the
centre of the domain and upward at the corners (not visible in this cut). This
causes the growth of the instability by increasing the sedimentation speed of
the DEM particles near the centre while reducing or even reversing the
sedimentation of those particles near the outer boundaries of the domain. The
movement of the DEM particles matches the expected behaviour of the instability.
Next we will attempt to quantitatively compare the SPH-DEM results to the growth
rate predicted by the analytical two-fluid model.

\begin{figure}
\centering
\includegraphics[height=0.9\textwidth,angle=-90]{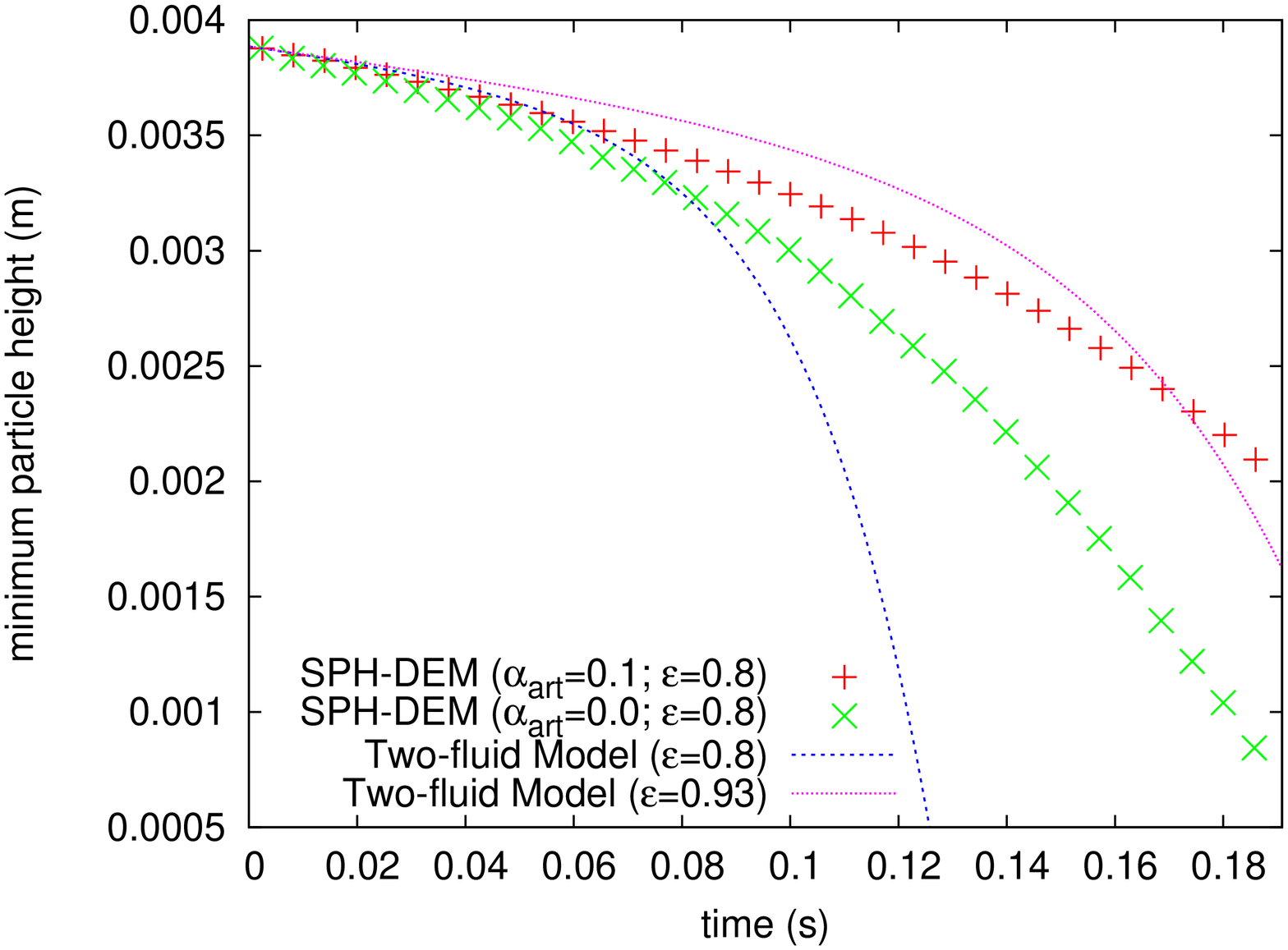}
\caption{Growth of the Rayleigh-Taylor instability using water. The red pluses
and green crosses show the position of the lowest DEM particle when the
artificial viscosity is either added or not. The two reference lines show the
growth rate predicted by a two-fluid model, using the lowest and highest
porosity of the CPB.}
\label{fig:multiGrowthRate_water}
\end{figure}

In Figure \ref{fig:multiGrowthRate_water} the growth of the RT instability
versus time for $\epsilon=0.8$, fluid resolution $h/d=2$ is shown using water as
the surrounding fluid. The symbols give the vertical position of the lowest DEM
particle, which provides an approximate measure of the instability amplitude
relative to an initially unperturbed situation. The vertical displacement of
this point over time can be compared with the estimated growth rate for the RT
instability as given by the two-fluid model in Eq.\ (\ref{eq:dispersionRT}). The
growth rate of the instability is added to the expected sedimentation speed
using Eq.\ (\ref{eq:expectedDiFeliceTermVel}) to calculate the expected
trajectory of the lowest DEM particle. Using the parameters of the simulation
and solving for the growth rate leads to a growth curve given by the lowest blue
dashed line. While a constant porosity of $0.8$ is used for the two-fluid RTI
model, the porosity of the DEM particles ranges from $0.8 \le \epsilon \le 0.86$
at $t=0$ (initial conditions) and the porosity at the leading front of the
instability grows over time, reaching a value of 0.93 at the time shown in
Figure \ref{fig:RTVis} and a maximum value of 0.95 before the instability meets
the bottom boundary. We use the analytical model to obtain an upper and lower
bound to the instability growth. The upper bound is calculated using
$\epsilon=0.8$ (the blue dashed line) and the lower bound (slower growth) is
calculated using $\epsilon=0.93$, which gives the purple dashed line. The
two-fluid model is included here as a benchmark, but it should be noted that
this model contains some significant approximations in treating the particle
suspension as an equivalent fluid, and is not necessarily more accurate than the
SPH-DEM results.

The SPH-DEM results are shown for the cases where the artificial viscosity is
either applied ($\alpha_{art}=0.1$) or not used ($\alpha_{art}=0.0$). In both
cases there is a clear exponential growth of the RT instability and only the
quantitative growth rate differs between the two simulations. Without the
artificial viscosity, the (exponential) growth rate lies between the two bounds.
After $t=0.15$ s the growth rate becomes slower than the upper bound, but by this time
the bottom of the instability is close to the bottom boundary, and we do not
expect the two-fluid model (which assumes small perturbations and an unbounded
domain) to apply. With artificial viscosity, the growth rate of
the instability is decreased and becomes slower than both of the two reference
bounds.

\begin{figure}
\centering
\includegraphics[angle=-90,width=0.9\textwidth]{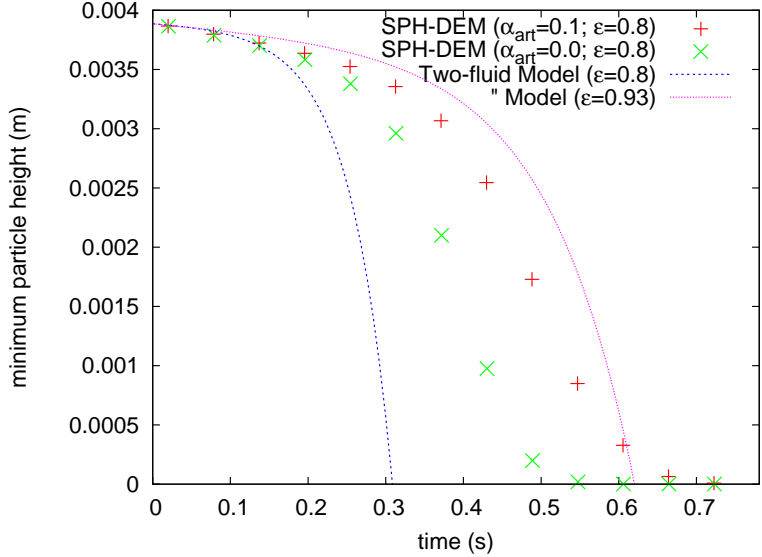}
\caption{Growth of the Rayleigh-Taylor instability using water-glycerol. The red
pluses and green crosses show the position of the lowest DEM particle when the
artificial viscosity is either added or not. The two reference lines show the
growth rate predicted by a two-fluid model, using the lowest and highest
porosity of the CPB.}
\label{fig:multiGrowthRate_water_glycerol}
\end{figure}

Figure \ref{fig:multiGrowthRate_water_glycerol} shows the same results but using
water-glycerol as the interstitial fluid. In this case the physical viscosity of
the fluid is proportionally greater than the artificial viscosity applied, and
therefore the addition of the artificial viscosity has a lesser effect. For both
$\alpha_{art}=0.1$ and $\alpha_{art}=0.0$ the growth rate of the instability
lies between the two bounds, except when the DEM particles reach the bottom of
the domain and large amplitude and wall effects dominate.

While it is encouraging that the SPH-DEM results closely match the expected
growth of the RT instability, the results highlight the negative effect of the
artificial viscosity when used in problems where the fluid or suspension
viscosity are important. It is therefore desirable to develop other approaches
to reduce the velocity fluctuations near high porosity gradients, and this is
the subject of current work. However, it is important to note that for the
majority of applications the addition of a small amount of artificial viscosity
has no significant effect on the results and is successful in eliminating the
problematic velocity fluctuations. For the interested reader, please see
\citet{colagrossi03numerical,gomez2010state,monaghan1994simulating} for a few
more examples where a similar SPH artificial viscosity has been successfully
applied.


In summary, the results from the RTI simulations using water-glycerol show that
the SPH-DEM simulation can accurately reproduce the Rayleigh-Taylor instability.
The addition of an artificial viscosity, while successful in dampening the
spurious velocity fluctuations, increases the effective viscosity of the system
and slightly reduces the growth rate of the instability.

\section{Conclusion}

We have presented a SPH implementation of the locally averaged Navier Stokes
equations and coupled this with a DEM model in order to provide a simulation
tool for two-way coupled fluid-particle systems. One notable property of
the resulting method is that it is completely particle-based and avoids the use
of a mesh. It is therefore suitable for those applications where a mesh
presents additional problems, for example, free surface flow or flow around
complex, moving and/or intermeshed geometries \citep{robinson2012dispersion}.

Furthermore, as the second main contribution of this study, we proposed a
validation procedure with test cases of increasing complexity (which can be
applied also to other methods).

The SPH-DEM formulation was used for 3D single and multiple particle
sedimentation problems and compared against analytical solutions for validation.

For single particle sedimentation (SPS) the simulations reproduced the
analytical solutions very well, with less than 1\% error over a wide range of
Particle Reynolds Numbers $0.011 \le Re_p \le 9$ and fluid resolutions. Only
when the fluid resolution became less than two times the particle diameter
did the results start to diverge from the expected solution.

For the multiple particle sedimentation test case using the Constant Porosity
Block (CPB), the SPH-DEM method accurately reproduced the expected terminal
velocity of the block within 5\% over prediction, over a range of porosities
$0.5< \epsilon <1.0$ and Particle Reynolds Numbers $0.002 \le Re_p \le 0.85$.
The over prediction of the terminal velocity is due to smoothing of the porosity
field near the edges of the block and reduces with a finer fluid resolution.
This error can be considered acceptable, considering the much lower
computational cost of SPH-DEM with the respect to the more accurate simulations
that can be obtained using a finely resolved FEM or Lattice Boltzmann method.

Further results from the CPB test case showed fluctuations in velocity of the
SPH particles near the edges of the block, which are likely due to fluctuations
in the pressure field being amplified by sudden changes in porosity. Adding a
small amount of artificial viscosity to the simulations was sufficient to damp
these fluctuations and prevent them from affecting the terminal velocity of the
block.

The Rayleigh-Taylor Instability (RTI) test case successfully reproduced the
instability and its growth rate for both water and water-glycerol. For this test
case the addition of artificial viscosity was not necessary for stability; due
to the relatively high porosity $\epsilon=0.8$ and lower porosity gradients at the
interface between the suspension and clear fluid.

Overall, the SPH-DEM model successfully reproduced the expected results
from the analytical test cases over a wide range of Reynolds Numbers and
porosities, and promises to be a flexible and accurate tool for modelling
particle-fluid systems. 

Current work is addressing the SPH velocity fluctuations near high porosity
gradients and promising results have already been obtained by either calculating
the drag separately on the fluid or re-deriving the SPH equations from a
Lagrangian formulation. In the future, the method will be applied to dispersion
of solids in liquid or liquid-gas environments \citep{robinson2012dispersion}.
Other relevant directions for future developments are: (i) the choice of
appropriate drag laws (e.g. for polydisperse flows) and the inclusion of the
added mass and lift forces; (ii) more realistic DEM particle contact forces and
(iii) the inclusion of contact friction and lubrication forces; and the
inclusion of surface tension effects.

\section*{Acknowledgment}
This work was supported by the PARDEM (www.pardem.eu) collaboration, which is a
EU Funded Framework 7, Marie Curie Initial Training Network.
Thanks to use of the cluster supported by the two
STW grants on ``A Numerical Wave Tank for Complex Wave and Current Interactions"
of Bokhove and Van der Vegt and on ``Polydispersed Granular Flows through
Inclined Channels" by Bokhove, Kuipers, Van der Vegt and Luding.

\bibliographystyle{model2-names}
\bibliography{fluidParticle,thesis}

\begin{thebibliography}{35}
\expandafter\ifx\csname natexlab\endcsname\relax\def\natexlab#1{#1}\fi
\expandafter\ifx\csname url\endcsname\relax
  \def\url#1{\texttt{#1}}\fi
\expandafter\ifx\csname urlprefix\endcsname\relax\def\urlprefix{URL }\fi
\providecommand{\eprint}[2][]{\url{#2}}
\providecommand{\bibinfo}[2]{#2}
\ifx\xfnm\relax \def\xfnm[#1]{\unskip,\space#1}\fi
\bibitem[{Anderson and Jackson(1967)}]{anderson67fluid}
\bibinfo{author}{Anderson, T.B.}, \bibinfo{author}{Jackson, R.},
  \bibinfo{year}{1967}.
\newblock \bibinfo{title}{Fluid mechanical description of fluidized beds.
  equations of motion}.
\newblock \bibinfo{journal}{Industrial \& Engineering Chemistry Fundamentals}
  \bibinfo{volume}{6}, \bibinfo{pages}{527--539}.
\bibitem[{Chandrasekhar(1961)}]{chandrasekhar61hydrodynamic}
\bibinfo{author}{Chandrasekhar, S.}, \bibinfo{year}{1961}.
\newblock \bibinfo{title}{{Hydrodynamic and hydromagnetic stability}}.
\newblock \bibinfo{publisher}{Dover Pubns}.
\bibitem[{Chu and Yu(2008)}]{chu08numerical}
\bibinfo{author}{Chu, K.}, \bibinfo{author}{Yu, A.}, \bibinfo{year}{2008}.
\newblock \bibinfo{title}{Numerical simulation of complex particle-fluid
  flows}.
\newblock \bibinfo{journal}{Powder Technology} \bibinfo{volume}{179},
  \bibinfo{pages}{104--114}.
\bibitem[{Cleary et~al.(2006)Cleary, Sinnott and Morrison}]{cleary06prediction}
\bibinfo{author}{Cleary, P.}, \bibinfo{author}{Sinnott, M.},
  \bibinfo{author}{Morrison, R.}, \bibinfo{year}{2006}.
\newblock \bibinfo{title}{{Prediction of slurry transport in SAG mills using
  SPH fluid flow in a dynamic DEM based porous media}}.
\newblock \bibinfo{journal}{Minerals engineering} \bibinfo{volume}{19},
  \bibinfo{pages}{1517--1527}.
\bibitem[{Colagrossi and Landrini(2003)}]{colagrossi03numerical}
\bibinfo{author}{Colagrossi, A.}, \bibinfo{author}{Landrini, M.},
  \bibinfo{year}{2003}.
\newblock \bibinfo{title}{Numerical simulation of interfacial flows by
  {S}moothed {P}article {H}ydrodynamics}.
\newblock \bibinfo{journal}{Journal of Computational Physics}
  \bibinfo{volume}{191}, \bibinfo{pages}{448--475}.
\bibitem[{Coulson and Richardson(1993)}]{coulson93chemical}
\bibinfo{author}{Coulson, J.}, \bibinfo{author}{Richardson, J.},
  \bibinfo{year}{1993}.
\newblock \bibinfo{title}{Chemical Engineering}. volume~\bibinfo{volume}{2}.
\newblock \bibinfo{edition}{4} edition.
\bibitem[{Dallavalle(1948)}]{dallavalle48micromeritics}
\bibinfo{author}{Dallavalle, J.}, \bibinfo{year}{1948}.
\newblock \bibinfo{title}{{Micromeritics: the technology of fine particles}}.
\newblock \bibinfo{publisher}{Pitman, New York}.
\bibitem[{Deen et~al.(2007)Deen, Van Sint~Annaland, Van Der~Hoef and
  Kuipers}]{deen07review}
\bibinfo{author}{Deen, N.}, \bibinfo{author}{Van Sint~Annaland, M.},
  \bibinfo{author}{Van Der~Hoef, M.}, \bibinfo{author}{Kuipers, J.},
  \bibinfo{year}{2007}.
\newblock \bibinfo{title}{Review of discrete particle modeling of fluidized
  beds}.
\newblock \bibinfo{journal}{Chemical Engineering Science} \bibinfo{volume}{62},
  \bibinfo{pages}{28--44}.
\bibitem[{Di~Felice(1994)}]{difelice94voidage}
\bibinfo{author}{Di~Felice, R.}, \bibinfo{year}{1994}.
\newblock \bibinfo{title}{The voidage function for fluid-particle interaction
  systems}.
\newblock \bibinfo{journal}{International Journal of Multiphase Flow}
  \bibinfo{volume}{20}, \bibinfo{pages}{153--159}.
\bibitem[{Ergun(1952)}]{ergun52fluid}
\bibinfo{author}{Ergun, S.}, \bibinfo{year}{1952}.
\newblock \bibinfo{title}{Fluid flow through packed columns}.
\newblock \bibinfo{journal}{Chemical Engineering and Processing}
  \bibinfo{volume}{48}, \bibinfo{pages}{89--94}.
\bibitem[{Fernandez et~al.(2011)Fernandez, Cleary, Sinnott and
  Morrison}]{fernandez11using}
\bibinfo{author}{Fernandez, J.}, \bibinfo{author}{Cleary, P.},
  \bibinfo{author}{Sinnott, M.}, \bibinfo{author}{Morrison, R.},
  \bibinfo{year}{2011}.
\newblock \bibinfo{title}{{Using SPH one-way coupled to DEM to model wet
  industrial banana screens}}.
\newblock \bibinfo{journal}{Minerals Engineering} \bibinfo{volume}{24},
  \bibinfo{pages}{741--753}.
\bibitem[{Gingold and Monaghan(1977)}]{gingold77smoothed}
\bibinfo{author}{Gingold, R.A.}, \bibinfo{author}{Monaghan, J.J.},
  \bibinfo{year}{1977}.
\newblock \bibinfo{title}{Smoothed particle hydrodynamic: theory and
  application to non-spherical stars}.
\newblock \bibinfo{journal}{Monthly Notices of the Royal Astronomical Society}
  \bibinfo{volume}{181}, \bibinfo{pages}{375--389}.
\bibitem[{Van~der Hoef et~al.(2005)Van~der Hoef, Beetstra and
  Kuipers}]{hoef05lattice}
\bibinfo{author}{Van~der Hoef, M.}, \bibinfo{author}{Beetstra, R.},
  \bibinfo{author}{Kuipers, J.}, \bibinfo{year}{2005}.
\newblock \bibinfo{title}{{Lattice-Boltzmann simulations of low-Reynolds-number
  flow past mono-and bidisperse arrays of spheres: results for the permeability
  and drag force}}.
\newblock \bibinfo{journal}{Journal of fluid mechanics} \bibinfo{volume}{528},
  \bibinfo{pages}{233--254}.
\bibitem[{Hoomans(1996)}]{hoomans96discrete}
\bibinfo{author}{Hoomans, B.}, \bibinfo{year}{1996}.
\newblock \bibinfo{title}{Discrete particle simulation of bubble and slug
  formation in a two-dimensional gas-fluidised bed: A hard-sphere approach}.
\newblock \bibinfo{journal}{Chemical Engineering Science} \bibinfo{volume}{51},
  \bibinfo{pages}{99--118}.
\bibitem[{Hoomans et~al.(2000)Hoomans, Kuipers and van
  Swaaij}]{hoomans00granular}
\bibinfo{author}{Hoomans, B.}, \bibinfo{author}{Kuipers, J.},
  \bibinfo{author}{van Swaaij, W.}, \bibinfo{year}{2000}.
\newblock \bibinfo{title}{Granular dynamics simulation of segregation phenomena
  in bubbling gas-fluidised beds}.
\newblock \bibinfo{journal}{Powder Technology} \bibinfo{volume}{109},
  \bibinfo{pages}{41--48}.
\bibitem[{Jiang et~al.(2007)Jiang, Oliveira and Sousa}]{jiang07mesoscale}
\bibinfo{author}{Jiang, F.}, \bibinfo{author}{Oliveira, M.},
  \bibinfo{author}{Sousa, A.}, \bibinfo{year}{2007}.
\newblock \bibinfo{title}{{Mesoscale SPH modeling of fluid flow in isotropic
  porous media}}.
\newblock \bibinfo{journal}{Computer Physics Communications}
  \bibinfo{volume}{176}, \bibinfo{pages}{471--480}.
\bibitem[{Krieger(1959)}]{krieger59mechanism}
\bibinfo{author}{Krieger, I.}, \bibinfo{year}{1959}.
\newblock \bibinfo{title}{{A mechanism for non Newtonian flow in suspensions of
  rigid spheres}}.
\newblock \bibinfo{journal}{Trans. Soc. Rheol.} \bibinfo{volume}{3},
  \bibinfo{pages}{137--152}.
\bibitem[{Li et~al.(2007)Li, Chu and Sheng}]{li07saturated}
\bibinfo{author}{Li, X.}, \bibinfo{author}{Chu, X.}, \bibinfo{author}{Sheng,
  D.}, \bibinfo{year}{2007}.
\newblock \bibinfo{title}{{A saturated discrete particle model and
  characteristic-based SPH method in granular materials}}.
\newblock \bibinfo{journal}{Int. J. Numer. Meth. Engng} \bibinfo{volume}{72},
  \bibinfo{pages}{858--882}.
\bibitem[{Lucy(1977)}]{lucy77numerical}
\bibinfo{author}{Lucy, L.B.}, \bibinfo{year}{1977}.
\newblock \bibinfo{title}{A numerical approach to testing the fission
  hypothesis}.
\newblock \bibinfo{journal}{The Astronomical Journal} \bibinfo{volume}{82},
  \bibinfo{pages}{1013--1924}.
\bibitem[{{Monaghan}(1997)}]{monaghan97SPHRiemannSolvers}
\bibinfo{author}{{Monaghan}, J.J.}, \bibinfo{year}{1997}.
\newblock \bibinfo{title}{{SPH and Riemann Solvers}}.
\newblock \bibinfo{journal}{Journal of Computational Physics}
  \bibinfo{volume}{136}, \bibinfo{pages}{298--307}.
\bibitem[{{Monaghan}(2005)}]{monaghan05SPH}
\bibinfo{author}{{Monaghan}, J.J.}, \bibinfo{year}{2005}.
\newblock \bibinfo{title}{{Smoothed particle hydrodynamics}}.
\newblock \bibinfo{journal}{Reports of Progress in Physics}
  \bibinfo{volume}{68}, \bibinfo{pages}{1703--1759}.
\bibitem[{Monaghan et~al.(2003)Monaghan, Kos and Issa}]{monaghan03fluid}
\bibinfo{author}{Monaghan, J.J.}, \bibinfo{author}{Kos, A.},
  \bibinfo{author}{Issa, N.}, \bibinfo{year}{2003}.
\newblock \bibinfo{title}{Fluid motion generated by impact}.
\newblock \bibinfo{journal}{Journal of waterway, port, coastal and ocean
  engineering} \bibinfo{volume}{129}, \bibinfo{pages}{250--259}.
\bibitem[{Pereira et~al.(2010)Pereira, Prakash and Cleary}]{pereira10sph}
\bibinfo{author}{Pereira, G.}, \bibinfo{author}{Prakash, M.},
  \bibinfo{author}{Cleary, P.}, \bibinfo{year}{2010}.
\newblock \bibinfo{title}{{SPH modelling of fluid at the grain level in a
  porous medium}}.
\newblock \bibinfo{journal}{Applied Mathematical Modelling} .
\bibitem[{Potapov(2001)}]{potapov01liquid}
\bibinfo{author}{Potapov, A.}, \bibinfo{year}{2001}.
\newblock \bibinfo{title}{Liquid solid flows using smoothed particle
  hydrodynamics and the discrete element method}.
\newblock \bibinfo{journal}{Powder Technology} \bibinfo{volume}{116},
  \bibinfo{pages}{204--213}.
\bibitem[{Price(2012)}]{price12smoothed}
\bibinfo{author}{Price, D.}, \bibinfo{year}{2012}.
\newblock \bibinfo{title}{Smoothed particle hydrodynamics and
  magnetohydrodynamics}.
\newblock \bibinfo{journal}{Journal of Computational Physics}
  \bibinfo{volume}{231}, \bibinfo{pages}{759--794}.
\bibitem[{{Robinson} et~al.(2012){Robinson}, {Luding} and
  {Ramaioli}}]{robinson2012dispersion}
\bibinfo{author}{{Robinson}, M.}, \bibinfo{author}{{Luding}, S.},
  \bibinfo{author}{{Ramaioli}, M.}, \bibinfo{year}{2012}.
\newblock \bibinfo{title}{{Simulations of grain dispersion by liquid injection
  using SPH-DEM}}.
\newblock \bibinfo{journal}{to be published} .
\bibitem[{Robinson and Monaghan(2011)}]{robinson11direct}
\bibinfo{author}{Robinson, M.}, \bibinfo{author}{Monaghan, J.},
  \bibinfo{year}{2011}.
\newblock \bibinfo{title}{Direct numerical simulation of decaying
  two-dimensional turbulence in a no-slip square box using smoothed particle
  hydrodynamics}.
\newblock \bibinfo{journal}{International Journal for Numerical Methods in
  Fluids} .
\bibitem[{Shepard(1968)}]{shepard682Dinterp}
\bibinfo{author}{Shepard, D.}, \bibinfo{year}{1968}.
\newblock \bibinfo{title}{A two-dimensional interpolation function for
  irregularly-spaced data}, in: \bibinfo{booktitle}{Proceedings of the 1968
  23rd ACM national conference}, \bibinfo{organization}{ACM}. pp.
  \bibinfo{pages}{517--524}.
\bibitem[{Tsuji et~al.(1993)Tsuji, Kawaguchi and Tanaka}]{tsuji93discrete}
\bibinfo{author}{Tsuji, Y.}, \bibinfo{author}{Kawaguchi, T.},
  \bibinfo{author}{Tanaka, T.}, \bibinfo{year}{1993}.
\newblock \bibinfo{title}{Discrete particle simulation of two-dimensional
  fluidized bed}.
\newblock \bibinfo{journal}{Powder Technology} \bibinfo{volume}{77},
  \bibinfo{pages}{79--87}.
\bibitem[{Wachmann et~al.(1998)Wachmann, Kalthoff, Schwarzer and
  Herrmann}]{wachmann98collective}
\bibinfo{author}{Wachmann, B.}, \bibinfo{author}{Kalthoff, W.},
  \bibinfo{author}{Schwarzer, S.}, \bibinfo{author}{Herrmann, H.},
  \bibinfo{year}{1998}.
\newblock \bibinfo{title}{Collective drag and sedimentation: comparison of
  simulation and experiment in two and three dimensions}.
\newblock \bibinfo{journal}{Granular Matter} \bibinfo{volume}{1},
  \bibinfo{pages}{75--82}.
\bibitem[{Wen and Yu(1966)}]{wen66mechanics}
\bibinfo{author}{Wen, C.}, \bibinfo{author}{Yu, Y.}, \bibinfo{year}{1966}.
\newblock \bibinfo{title}{Mechanics of fluidization}.
\newblock \bibinfo{journal}{Chemical Engineering Programming Symposium Series}
  \bibinfo{volume}{62}, \bibinfo{pages}{100--108}.
\bibitem[{Xu(1997)}]{xu97numerical}
\bibinfo{author}{Xu, B.}, \bibinfo{year}{1997}.
\newblock \bibinfo{title}{Numerical simulation of the gas solid flow in a
  fluidized bed by combining discrete particle method with computational fluid
  dynamics}.
\newblock \bibinfo{journal}{Chemical Engineering Science} \bibinfo{volume}{52},
  \bibinfo{pages}{2785--2809}.
\bibitem[{Xu(2000)}]{xu00numerical}
\bibinfo{author}{Xu, B.}, \bibinfo{year}{2000}.
\newblock \bibinfo{title}{Numerical simulation of the gas solid flow in a bed
  with lateral gas blasting}.
\newblock \bibinfo{journal}{Powder Technology} \bibinfo{volume}{109},
  \bibinfo{pages}{13--26}.
\bibitem[{Xu et~al.(2007)Xu, Ge and Li}]{xu07discrete}
\bibinfo{author}{Xu, M.}, \bibinfo{author}{Ge, W.}, \bibinfo{author}{Li, J.},
  \bibinfo{year}{2007}.
\newblock \bibinfo{title}{A discrete particle model for particle-fluid flow
  with considerations of sub-grid structures}.
\newblock \bibinfo{journal}{Chemical engineering science} \bibinfo{volume}{62},
  \bibinfo{pages}{2302--2308}.
\bibitem[{Zhu et~al.(1999)Zhu, Fox and Morris}]{zhu99pore}
\bibinfo{author}{Zhu, Y.}, \bibinfo{author}{Fox, P.}, \bibinfo{author}{Morris,
  J.}, \bibinfo{year}{1999}.
\newblock \bibinfo{title}{{A pore-scale numerical model for flow through porous
  media}}.
\newblock \bibinfo{journal}{International journal for numerical and analytical
  methods in geomechanics} \bibinfo{volume}{23}, \bibinfo{pages}{881--904}.

\end{thebibliography}

\end{document}